\documentclass{elsart}

\usepackage{epsfig}

\usepackage{amssymb}

\def\pat{\partial}

\def \ind{\indent{~~~}}
\begin{document}

\begin{frontmatter}



\title{An Analytical Study on the Instability Phenomena During the Phase Transitions in a Thin Strip under
Uniaxial Tension }


\author[a]{Hui-Hui Dai},
\ead{ mahhdai@cityu.edu.hk }
\author[b]{Zongxi Cai}

\address[a]{Department of Mathematics and Liu
Bie Ju Centre for Mathematical Sciences, City University of Hong
Kong, 83 Tat Chee Avenue, Hong Kong, P.R. China}
 \address[b]{Department of Mechanics, Tianjin University, Tianjin, 300072, P.R.
China}

\begin{abstract}
\ind In the experiments on stress-induced phase transitions in
 SMA strips, several interesting instability phenomena have been observed, including
 a necking-type instability (associated with the stress drop), a shear-type instability
 (associated with the inclination of the transformation front) and an orientation instability
 (associated with the switch of the inclination angle). In order to shed more light on these
 phenomena, in this paper we conduct an analytical study. We
 consider the problem in a three-dimensional setting, which
 implies that one needs to study the difficult problem of solution
 bifurcations of high-dimensional nonlinear partial differential
 equations. By using the smallness of the maximum strain, the thickness and
 width of the strip, we use a methodology, which combines series
 expansions and asymptotic expansions, to derive the asymptotic
 normal form equations, which can yield the leading-order behavior
 of the original three-dimensional field equations. An important feature of the
 second normal form equation is that it contains a turning point for the localization (necking) solution
of the first equation. It is the presence of such a turning point
which causes the inclination of the phase front. The WKB method is
used to construct the asymptotic solutions, which can capture the
shear instability and the orientation instability successfully.
Our analytical results reveal that the inclination of the phase
front is a phenomenon of localization-induced buckling (or
phase-transition-induced buckling as the localization is caused by
the phase transition). Due to the similarities between the
development of the Luders band in a mild steel and the
stress-induced transformations in a SMA, the present results give
a strong analytical evidence that the former is also caused by
macroscopic effects instead of microscopic effects. Our analytical
results also reveal more explicitly the important roles played by
the geometrical parameters.

\end{abstract}

\end{frontmatter}

% main text
\section{Introduction}
\label{s1}

\ind Shape memory alloys (SMAs), due to their two important
characteristics, shape memory effect and pseudoelasticity, have
broad applications (see Duerig et al 1990, Otsuka and Wayman
1998). To understand the behaviour of this type of materials,
systematic experiments have been carried out on uniaxial tension
of superelastic NiTi (a kind of SMAs) wires, strips and tubes (see
Shaw \& Kyriakides 1995, 1997, 1998, Sun et al 2000, Tse \& sun
2000, Feng \& Sun 2006, Chang et al 2006). An experimental vedio
on the tension of a NiTi  strip by Q. P. Sun's group can be found
in the website: www.me.ust.hk/\~meqpsun/vedio/tension-strip.htm.
Among many important observations in these experiments, some key
features are the various instability phenomena associated with
stress-induced phase transitions. For example, for the
stress-induced phase transitions in a strip during a loading
process, at least three instability phenomena were observed (see
Shaw \& Kyriakides 1998, Sun et al 2000, Tse \& Sun 2000): (i) a
stress drop after the nucleation of the martensite phase and
associated with it there is a formation of two phase fronts, which
manifest like a neck (a necking-type instability); (ii) the phase
front inclines an angle with the strip axis (a shear-type
instability); (iii) the front can switch the inclination to an
opposite angle. Finite element simulations have been carried out
(see Shaw \& Kyriakides 1998, Shaw 2000) to capture the main
features observed in experiments. The numerical results in these
two papers revealed some important information about the
stress-induced phase transitions in strips. For example, it was
found that the evolution of phase transition events is strongly
influenced by overall geometric (structural) effects. The results
of these two papers strongly suggest that continuum level events
remain dominant players in the SMAs considered by them.

\ind Motivated by the experimental and numerical results by others
mentioned above, in this paper we shall study instability
phenomena during the phase transitions in a strip {\it
analytically}. We model this problem in a continuum
three-dimensional setting with a no-convex strain energy function,
in view of the results of Shaw \& Kyriakides (1998) (cf. the last
sentence of the paragraph above). The difference between the
stress-strain relation used in this paper and Shaw \& Kyriakides
(1998) is that the former is a cubic nonlinear curve while the
latter is a trilinear one. The nonlinearity could play certain
role (see, Fig. 8 of Shaw 2000). Since in the experiments, the
maximum strain is less than $8\%$, keeping the nonlinearity up to
the third order (see (2.3)) is accurate enough, at least not worse
than a trilinear approximation. We also point out that here the
intention is to study macroscopic instability phenomena in the
loading process only and no attempt is made to consider the
microscopic effects.

\ind Analytical results, if achievable, have a number of
advantages. One is that there is no need to introduce an
artificial imperfection to capture the post-bifurcation mode.
Secondly, mathematically, an instability is caused by the fact
that there are multiple solutions, and analytical results can shed
light on how this situation arises and help to understand the
mechanism. Thirdly, from the analytical results one can see more
clearly the roles played by various parameters (for the present
problem, in particular, the geometric parameters, e.g. the
thickness and the width). Indeed, the analytical results obtained
in this paper reveal more explicitly the important role of the
thickness of the strip and show that the width influences the
instability phenomena through the thickness-width ratio rather its
magnitude.

\ind As pointed out in Shaw \& Kyriakides (1998), in the
macroscopic scale there are many similarities between
stress-induced transformations in a SMA and the development of
Luders bands in a mild steel. In the literature, there are
different views whether the Luders band is caused by microscopic
effects or macroscopic effects (see Estrin \& Kubin 1995). Here,
we have shown that the inclination of the transformation front is
a phenomenon of localization-induced buckling. This offers a
strong analytical evidence that this phenomenon in a SMA, and
plausibly the phenomenon of Luders bands (due to the
similarities), is due to macroscopic effects.

\ind Since we formulate this problem in a three-dimensional
setting with a nonlinear constitutive relation, the governing
field equations are three coupled nonlinear partial differential
equations (PDEs). It is extremely difficult to deduce the
post-bifurcation solutions of nonlinear PDEs analytically.
Fortunately, for the present problem, several quantities are
small, e.g., the thickness, the width, and the maximum strain,
which then permit us to use a methodology of coupled series and
asymptotic expansions to deduce asymptotic solutions. This
methodology was first introduced to study nonlinear waves in
solids (see Dai \& Huo 2002, Dai \& Fan 2004). Recently, it has
been successfully used to study various instability phenomena in
solids (see Dai \& Cai 2006, Cai \& Dai 2006, Dai et al 2008, Dai
\& Wang 2008). However, all those problems studied before are
essentially two-dimensional. Here, for the first time this
methodology is used to study a three-dimensional problem.

\ind The remaining of this paper is arranged as follows. In
section 2,  we give the general three-dimensional field equations
for a plate. Then, in section 3, we non-dimensionalize the
three-dimensional governing equations to identify the key small
variables and small parameters for a thin plate. And then, by
using the smallness of these variables and parameters and a
methodology of coupled series and asymptotic expansions, in
section 4 we derive the asymptotic two-dimensional equations. By
considering the smallness of the width further and using a similar
methodology in section 4, we obtain the quasi one-dimensional
asymptotic normal form equations for a thin strip in section 5.
These normal form equations are then solved analytically in
section 6 for an infinitely long strip and the solutions obtained
seem to be able to describe many features observed in experiments.
Finally, some conclusions are drawn.

\section{Three-Dimensional Field  Equations}
\setcounter{equation}{0}

\ind We consider the deformation of a three-dimensional plate
composed of a hyperelastic material. Its thickness is $2a$ and its
width is $2b$. We use the Cartesian coordinates $(x,y,z)$
(equivalently $x_i$) and $(X,Y,Z)$ (equivalently $X_i$) to
represent a material point in the current and reference
configurations, respectively. The geometry of the object of study
is shown in Figure 1.
\begin{figure}[htb]
\includegraphics{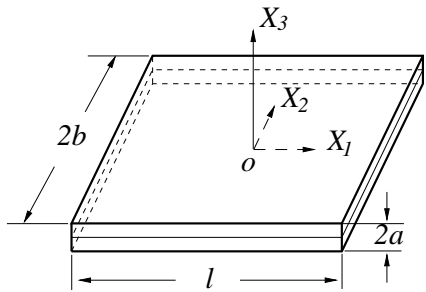} \vspace{6cm} \caption{The geometry of the object of
study }\label{fig1}
\end{figure}

\ind To ensure phase transitions can take place, we suppose that
the strain energy function $\Phi$, which is a function of the
invariants of the left Cauchy-Green strain tensor for a
homogeneous isotropic hyperelastic material, is non-convex such
that there is a local maximum and a local minimum in the uniaxial
stress-strain curve under a homogenous constant strain state. The
first Piola-Kirchhoff stress tensor {\boldmath{$\Sigma$}} is given
by
\begin{eqnarray}
{\bf{\Sigma}}={\frac {\pat \Phi}{\pat {\bf F}}},\label{e2.1}
\end{eqnarray}
where ${\bf F}$ is the deformation gradient and the  components of
$({\bf F}-{\bf I})$ are
\begin{eqnarray}
u_{i,j}={\frac {\pat u_i}{\pat X_j}}, ~~~~(i,j=1,2,3),\label{e2.2}
\end{eqnarray}
where $u_i$  are the components of the displacement vector and
$X_j$ are coordinates in the undeformed configuration. If the
strains are small, it is possible to expand the first
Piola-Kirchhoff stress components in term of the strains up to any
order. Due to the complexity of the problem, we shall consider the
material nonlinearity up to the third order. The formula
containing terms up to the third-order material nonlinearity has
been provided in Fu and Ogden (1999) as
\begin{eqnarray}
&&\Sigma_{ji}=a_{jilk}^1 u_{k,l} +{\frac {1}{2}}a_{jilknm}^2
u_{k,l} u_{m,n}+{\frac {1}{6}}a_{jilknmqp}^3 u_{k,l}u_{m,n}
u_{p,q},\label{e2.3}
\end{eqnarray}
where  $a_{jilk}^1$, $a_{jilknm}^2$ and $a_{jilknmqp}^3$ are
incremental elastic moduli, which can be calculated once a
specific form of the strain energy function is given. Their
expressions can be found in Appendix A, where it is also shown
that $a_{jilk}^1$ has 4 non-zero members and only two are
independent, $a_{jilknm}^2$ has 9 non-zero members and only three
are independent and $a_{jilknmqp}^3$ has 22 non-zero members and
only four are independent. For the convenience of the sequel
analysis, we write out the index $3$ explicitly. As a result, from
Eqs. (\ref{e2.3}) we have
\begin{eqnarray}
\Sigma_{ji}&=&a_{ji33}^1u_{3,3}+a_{ji\alpha 3}^1u_{3,\alpha}
+a_{ji3\alpha}^1u_{\alpha,3}+a_{ji\beta\alpha}^1u_{\alpha,\beta}\nonumber\\
&+&{\frac {1}{2}}( a_{ji3333}^2u_{3,3}^2+2(a_{ji\alpha
333}^2u_{3,\alpha} +a_{ji3\alpha 33}^2u_{\alpha,3}+a_{ji\beta
\alpha 33}^2u_{\alpha,\beta})
u_{3,3}\nonumber\\
&+&a_{ji\alpha 3\beta 3}^2u_{3,\alpha}u_{3,\beta}+a_{ji3\alpha
3\beta}^2u_{\alpha,3}u_{\beta,3}+a_{ji\beta \alpha \sigma
\gamma}^2u_{\alpha,\beta}u_{\gamma,\sigma}\nonumber\\
&+&2(a_{ji3\beta \alpha 3}^2u_{3,\alpha}u_{\beta,3} +a_{ji\beta
\alpha \gamma 3}^2u_{\alpha,\beta}u_{3,\gamma} +a_{ji\beta \alpha
3\gamma}^2u_{\alpha,\beta}u_{\gamma,3}))\nonumber\\
&+&{\frac {1}{6}}(3(a_{ji\beta\alpha
3333}^3u_{\alpha,\beta}+a_{ji\alpha
33333}^3u_{3,\alpha}+a_{ji3\alpha
3333}^3u_{\alpha,3})u_{3,3}^2\nonumber\\
&+&3(a_{ji\alpha 3\beta 333}^3u_{3,\alpha}u_{3,\beta}+a_{ji3\alpha
3\beta 33}^3u_{\alpha,3}u_{\beta,3}+a_{ji\beta\alpha\sigma\gamma
33}^3u_{\alpha,\beta}u_{\gamma,\sigma})u_{3,3}\nonumber\\
&+&6(a_{ji3\beta\alpha
333}^3u_{3,\alpha}u_{\beta,3}+a_{ji\beta\alpha 3\gamma
33}^3u_{\alpha,\beta}u_{\gamma,3}+a_{ji\beta\alpha\sigma
333}^3u_{\alpha,\beta}u_{3,\gamma})u_{3,3}\nonumber\\
&+&a_{ji\alpha 3\beta 3\gamma
3}^3u_{3,\alpha}u_{3,\beta}u_{3,\gamma}+(a_{ji3\alpha 3\beta
3\gamma}^3u_{\alpha,3}u_{\beta,3}+3a_{ji\alpha 3\beta
33\gamma}^3u_{3,\alpha}u_{3,\beta}\nonumber\\
&+&3a_{ji\alpha 33\beta
3\gamma}^3u_{3,\alpha}u_{\beta,3})u_{\gamma,3}+(3a_{ji\alpha
3\beta 3\sigma\gamma}^3u_{3,\alpha}u_{3,\beta}+3a_{ji3\alpha
3\beta
\sigma\gamma}^3u_{\alpha,3}u_{\beta,3}\nonumber\\
&+&6a_{ji3\alpha \beta
3\sigma\gamma}^3u_{\alpha,3}u_{3,\beta}+3a_{ji3\tau
\beta\alpha\sigma\gamma}^3u_{\tau,3}u_{\alpha,\beta} +3a_{ji\tau
3\beta\alpha\sigma\gamma}^3u_{3,\tau}u_{\alpha,\beta})u_{\gamma,\sigma}
\nonumber\\
&+&a_{ji333333}^3u_{3,3}^3 +a_{ji\tau\lambda
\beta\alpha\sigma\gamma}^3u_{\lambda,\tau}u_{\alpha,\beta}u_{\gamma,\sigma})
,~~~(\alpha,
\beta,\gamma,\sigma,\tau,\lambda=1,2).\nonumber\\&&\label{e2.4}
\end{eqnarray}
The equations of equilibrium are
\begin{eqnarray}
\Sigma_{ji,j}=0.\label{e2.5}
\end{eqnarray}
Substituting (\ref{e2.4}) into (\ref{e2.5}), we obtain
\begin{eqnarray}
&&b_{i333}u_{3,33}+(b_{i3\alpha3}+b_{i\alpha33})u_{3,3\alpha}
+b_{i\beta\alpha 3}u_{3,\alpha \beta}\nonumber\\
&+&b_{i33\alpha}u_{\alpha,33}+(b_{i\beta3\alpha}+b_{i
3\beta\alpha})u_{\alpha,3\beta}
+b_{i\gamma\beta\alpha}u_{\alpha,\beta\gamma}=0,\label{e2.6}
\end{eqnarray}
where
\begin{eqnarray}
b_{ijkl}&=&a_{jilk}^1+a_{jilk33}^2u_{3,3}+a_{jilk\beta\alpha}^2u_{\alpha,\beta}
+a_{jilk3\alpha}^2u_{\alpha,3}+a_{jilk\alpha3}^2u_{3,\alpha}\nonumber\\
&+&{\frac {1}{2}} a_{jilk3333}^3u_{3,3}^2 +(a_{jilk\beta \alpha
33}^3u_{\alpha,\beta}+a_{jilk\alpha 333}^3u_{3,\alpha}
+a_{jilk3\alpha 33}^3u_{\alpha,3})u_{3,3}\nonumber\\
&+&{\frac {1}{2}}a_{jilk\beta \alpha \sigma
\gamma}^3u_{\alpha,\beta}u_{\gamma,\sigma} +(a_{jilk\alpha 3\sigma
\gamma}^3u_{3,\alpha} +a_{jilk3\alpha \sigma
\gamma}^3u_{\alpha,3})u_{\gamma,\sigma}\nonumber\\
&+&a_{jilk\beta 33\alpha }^3u_{\alpha,3}u_{3,\beta} +{\frac
{1}{2}}(a_{jilk\beta 3\alpha
3}^3u_{3,\alpha}u_{3,\beta}+a_{jilk3\beta3\alpha }^3u_{\alpha,3}
u_{\beta,3}).\label{e2.7}
\end{eqnarray}
It should be noted that, for an initially isotropic material these
terms in the right hand side of (\ref{e2.7}) will vanish when the
number of index $3$ in their all subscripts is odd.

\ind Eqs. (\ref{e2.6}) are the governing equations for the three
unknowns $u_i (i=1,2,3)$. In order to investigate the instability
phenomena, one needs to study the solution bifurcations of the
three-dimensional nonlinear partial differential equations (PDE's)
(\ref{e2.6}) with the traction free conditions on the top/bottom
surfaces and two side surfaces and under proper end conditions.
Mathematically, this is a very challenging problem, since there is
no available a general method for studying the bifurcations of
three-dimensional nonlinear PDE's. Here, we shall use a novel
approach involving coupled series and asymptotic expansions to
derive the asymptotic normal form equations in order to carry out
the analysis. For that purpose, we first non-dimensionalize the
governing equations to identify the key small variables and small
parameters.

\section{Non-dimensional  Equations}
\setcounter{equation}{0}

\ind Suppose that the loads acting on the boundaries of the plate
are symmetrical about the mid plane and therefore the deformation
is also symmetrical about this plane. Then, we have
\begin{eqnarray}
u_{\alpha}(X_{\beta},-X_3)=u_{\alpha}(X_{\beta},X_3),~~~~
u_3(X_{\beta},-X_3)=-u_3(X_{\beta},X_3).\label{e3.1}
\end{eqnarray}
Based on Eq. (\ref{e3.1}), we introduce a transformation
\begin{eqnarray}
u_3=X_3 w,~~~~ s=X_3^2.\label{e3.2}
\end{eqnarray}
The dimensionless quantities are defined through the following
scalings:
\begin{eqnarray}
s=l^2\tilde s,~~~ X_{\alpha}=l\tilde x_{\alpha}, ~~~
u_{\alpha}=h\tilde u_{\alpha},~~~w={\frac {h}{l}}\tilde
w,\label{e3.3}
\end{eqnarray}
where $l$ is the length of the plate and $h$ is a characteristic
displacement in the mid-plane. From (\ref{e3.2}) and (\ref{e3.3}),
we obtain
\begin{eqnarray}
&&u_{3,3}=\epsilon (\tilde w+2\tilde s {\frac {\pat \tilde w}{\pat
\tilde s}}),~~~~u_{\alpha,\beta}=\epsilon  {\frac {\pat \tilde
u_\alpha}{\pat \tilde x_\beta}},\nonumber\\
&&u_{3,\alpha}=\epsilon \sqrt{\tilde s} {\frac {\pat \tilde
w}{\pat \tilde x_\alpha}},~~~~u_{\alpha,3}=2\epsilon \sqrt{\tilde
s} {\frac {\pat \tilde u_\alpha}{\pat \tilde s}},\label{e3.4}
\end{eqnarray}
where $\epsilon=h/l$ is a small parameter (i.e., we are
considering a weak nonlinearity). The second-order derivatives can
be treated similarly.

\ind Substituting (\ref{e3.3}) and (\ref{e3.4}) into (\ref{e2.7})
and (\ref{e2.6}), we obtain
\begin{eqnarray}
&&a_{\beta\tau\gamma\alpha}^1u_{\alpha,\beta\gamma}+2a_{3\tau
3\alpha}^1u_{\alpha,s}
+(a_{3\tau\alpha3}^1+a_{\alpha\tau33}^1)w_{,\alpha}\nonumber\\
&+&s(4a_{3\tau 3\alpha}^1u_{\alpha,ss}+2(a_{3\tau\alpha3}^1
+a_{\alpha\tau33}^1)w_{,\alpha s})\nonumber\\
&+&\epsilon
(a_{\beta\tau\gamma\alpha\eta\xi}^2u_{\alpha,\beta\gamma}u_{\xi,\eta}
+2a_{3\tau3\alpha\eta\xi}^2u_{\alpha,s}u_{\xi,\eta}
+a_{\beta\tau\gamma\alpha33}^2u_{\alpha,\beta\gamma}w
+2a_{3\tau\alpha333}^2u_{\alpha,s}w\nonumber\\
&&+(a_{3\tau\alpha3\eta\xi}^2+a_{\alpha\tau33\eta\xi}^2)u_{\xi,\eta}w_{,\alpha}
+(a_{3\tau\alpha333}^2+a_{\alpha\tau3333}^2)ww_{,\alpha}\nonumber\\
&+&s(4(a_{3\tau\alpha\beta\eta3}^2+a_{\beta\tau3\alpha\eta3}^2)u_{\alpha,\beta
s}u_{\eta,s}+4a_{3\tau\alpha3\beta\eta}^2u_{\alpha,s
s}u_{\eta,\beta}\nonumber\\
&&+4a_{3\tau\alpha333}^2(u_{\alpha,s s}w+u_{\alpha,s}w_{,s})
+2a_{\alpha\tau\beta3\eta3}^2u_{\eta,s}w_{,\alpha\beta}
+2a_{\beta\tau\gamma\alpha33}^2u_{\alpha,\beta\gamma}w_{,s}\nonumber\\
&&+12a_{3\tau33\alpha3}^2u_{\alpha,s}w_{,s}
+2(a_{3\tau\alpha333}^2+a_{\alpha\tau3333}^2
+3a_{3\tau333\alpha}^2)w_{,\alpha}w_{,s}\nonumber\\
&&+2(a_{3\tau3\gamma\beta\alpha}^2
+a_{\gamma\tau33\beta\alpha}^2)u_{\alpha,\beta}w_{,\gamma s}
+2(a_{\gamma\tau3333}^2+a_{3\tau\gamma333}^2)ww_{,\gamma s}\nonumber\\
&&+2(a_{3\tau\beta\alpha3\eta}^2+a_{\beta\tau3\alpha3\eta}^2)u_{\alpha,\beta
s}w_{,\eta}+a_{\alpha\tau\beta33\eta}^2w_{,\alpha\beta}w_{,\eta})\nonumber\\
&+&4s^2(2a_{3\tau\alpha333}^2u_{\alpha,s s}w_{,s}
+(a_{\alpha\tau3333}^2+a_{3\tau\alpha333}^2)w_{\alpha,s}w_{,s}\nonumber\\
&&+2a_{3\tau33\alpha3}^2u_{\alpha,s}w_{,ss}
+a_{3\tau333\alpha}^2w_{,\alpha}w_{,ss}))\nonumber\\
&+&\epsilon^2 (H_1)=0~~~~~~(\tau=1,2),\label{e3.5}
\end{eqnarray}
\begin{eqnarray}
&&2(a_{33\beta\alpha}^1+a_{\beta3 3\alpha}^1)u_{\alpha,\beta
s}+a_{\alpha3\beta3}^1w_{,\alpha\beta}
+6a_{3333}^1w_{,s}+4sa_{3333}^1w_{,ss}\nonumber\\
&+&\epsilon
(2a_{\beta3\gamma\alpha\eta3}^2u_{\alpha,\beta\gamma}u_{\eta,s}
+4a_{\alpha333\eta3}^2u_{\alpha,s}u_{\eta,s}
+2(a_{33\beta\alpha\eta\xi}^2+a_{\alpha3\beta3\eta\xi}^2)u_{\alpha,\beta
s}u_{\xi,\eta}\nonumber\\
&&+2(a_{33\beta\alpha33}^2+a_{\alpha3\beta333}^2)u_{\alpha,\beta
s}w+2(a_{\eta3333\beta}^2+a_{\beta333\eta3}^2)u_{\eta,s}w_{,\alpha}\nonumber\\
&&+a_{\alpha3\beta3\eta\xi}^2u_{\xi,\eta}w_{,\alpha\beta}
+a_{\alpha3\beta333}^2ww_{,\alpha\beta}
+6a_{3333\beta\alpha}^2u_{\alpha,\beta}w_{,s}
+6a_{333333}^2ww_{,s}\nonumber\\
&&+a_{\beta3\gamma\alpha3\eta}^2u_{\alpha,\beta\gamma}w_{,\eta}
+2a_{\alpha3333\eta}^2u_{\alpha,s}w_{,\eta}
+2a_{\alpha3333\beta}^2w_{,\alpha}w_{,\beta}\nonumber\\
&+&2s (4a_{\alpha333\beta3}^2u_{\alpha,ss}u_{\beta,s}
+2(a_{33\beta\alpha33}^2+a_{\alpha3\beta333}^2)u_{\alpha,\beta
s}w_{,s}+a_{\alpha3\beta333}^2w_{,\alpha\beta}w_{,s}\nonumber\\
&&+a_{333333}^2(2ww_{,ss}+6w_{,s}^2)
+4a_{\alpha333\beta3}^2u_{\beta,s}w_{,\alpha
s}+2a_{3333\beta\alpha}^2u_{\alpha,\beta}w_{,ss}\nonumber\\
&&+2a_{\alpha3333\beta}^2u_{\alpha,ss}w_{,\beta}
+(a_{\alpha3333\beta}^2+a_{\alpha3333\beta}^2)w_{,\beta}w_{,\alpha
s})\nonumber\\
&+&8s^2a_{333333}w_{,s}w_{,ss})\nonumber\\
&+&\epsilon^2 (H_2)=0,\label{e3.6}
\end{eqnarray}
where and thereafter the tilde over non-dimensional variables has
been dropped for convenience. The lengthy expressions for
$H_i(i=1,2,\cdots$) are omitted although they are needed for the
calculations (interested readers can contact the corresponding
author for their expressions). We consider the case that the top
and bottom surfaces of the plate are traction-free. By using
(\ref{e2.4}), we have
\begin{eqnarray}
&&2a_{3\tau\alpha3}^1u_{\alpha,s}+a_{3\tau3\alpha}^1w_{,\alpha}\nonumber\\
&+& {\epsilon}
(4a_{3\tau\alpha\beta\gamma3}^2u_{\beta,\alpha}u_{\gamma,s}
+4a_{3\tau\alpha333}^2u_{\alpha,s}w
+2a_{3\tau\alpha\beta3\gamma}^2u_{\beta,\alpha}w_{,\gamma}
+2a_{3\tau3\alpha33}^2ww_{,\beta}\nonumber\\
&+&2s w_{,s}(2a_{3\tau\alpha333}^2u_{\alpha,s}
+a_{3\tau3\alpha33}^2w_{,\alpha}))\nonumber\\
&+& {\epsilon^2}(H_3)|_{s=\nu_1}=0~~~~~~(\tau=1,2),\label{e3.7} \\
&&\nonumber\\
&&a_{33\alpha\beta}^1u_{\beta,\alpha}+a_{3333}^1(w+2s w_{,s})\nonumber\\
&+&{\frac {\epsilon}{2}}
(a_{33\alpha\beta\gamma\delta}^2u_{\beta,\alpha}u_{\delta,\gamma}
+2a_{33\alpha\beta33}^2u_{\beta,\alpha}w
+a_{333333}^2w^2\nonumber\\
&+&s(4a_{33\alpha3\beta3}^2u_{\alpha,s}u_{\beta,s}
+4a_{33\alpha\beta33}^2u_{\beta,\alpha}w_{,s}+4a_{333333}^2ww_{,s}\nonumber\\
&&+4a_{33\alpha33\beta}^2u_{\alpha,s}w_{,\beta}
+a_{333\alpha3\beta}^2w_{,\alpha}w_{,\beta})+4s^2 a_{333333}^2w_{,s}^2)\nonumber\\
&+&{\epsilon^2} (H_4)|_{s=\nu_1}=0,\label{e3.9}
\end{eqnarray}
where $\nu_1=a^2/l^2$ . Please note that due to symmetry, the
boundary conditions at the bottom surface are automatically
satisfied.

\ind Eqs. (\ref{e3.5}) and (\ref{e3.6}) provide the governing
equations for three unknowns $u_\alpha$ and $w$ and the boundary
conditions are (\ref{e3.7}) and (\ref{e3.9}). However, they still
comprise a formidable system of nonlinear PDE's to be analyzed
directly. To go further, we assume that the plate is thin. Then
$\nu_1$ is a small parameter and $0\le s\le \nu_1$ is a small
variable. It is clear that the unknowns are functions of the
spatial variables $x_1$ and $x_2$, the small variable $s$ and two
small parameters $\nu_1$ and $\epsilon$. Next, we shall use the
smallness of the variable $s$ and two parameters $\nu_1$ and
$\epsilon$ to derive the asymptotic two-dimensional equations.

{\bf Remark:} Since the current methodology to deduce the
one-dimensional asymptotic normal equations from the
three-dimensional nonlinear field equations has not been done
before, in the next two sections we shall provide some detailed
derivations.

\section{Two-dimensional Asymptotic Equations}
\setcounter{equation}{0}

\ind As discussed in the previous section, we can write
\begin{eqnarray}
u_{\alpha}=u_{\alpha}(x_{\beta},s;\epsilon,\nu_1),~~~~
w=w(x_{\beta},s;\epsilon,\nu_1).\label{e4.1}
\end{eqnarray}
As the variable $s$ is small, as long as we assume that the
unknowns are sufficiently smooth in $s$, we can take the series
expansions in $s$ for the unknowns, i.e.,
\begin{eqnarray}
u_{\alpha}&=&U_{0\alpha}(x_{\beta};\epsilon,\nu_1)+sU_{1\alpha}(x_{\beta};\epsilon,\nu_1)
+s^2U_{2\alpha}(x_{\beta};\epsilon,\nu_1)+\cdots,\nonumber\\
w&=&W_{0}(x_{\beta};\epsilon,\nu_1)+sW_{1}(x_{\beta};\epsilon,\nu_1)
+s^2W_{2}(x_{\beta};\epsilon,\nu_1)+\cdots .\label{e4.2}
\end{eqnarray}
Substituting Eq. (\ref{e4.2}) into the boundary conditions
(\ref{e3.7}) and (\ref{e3.9}), and noting that $s=\nu_1$, we
obtain
\begin{eqnarray}
&&2a_{3\tau\alpha3}^1U_{1\alpha}+a_{3\tau3\alpha}^1W_{0,\alpha}
+\nu_1(4a_{3\tau\alpha3}^1U_{2\alpha}+a_{3\tau3\alpha}^1W_{1,\alpha})\nonumber\\
&+&{\frac
{\epsilon}{2}}(4a_{3\tau\alpha3\eta\xi}^2U_{0\xi,\eta}U_{1\alpha}
+4a_{3\tau\alpha333}^2W_0U_{1\alpha}
\nonumber\\
&&+2a_{3\tau3\alpha\eta\xi}^2U_{0\xi,\eta}W_{0,\alpha}
+2a_{3\tau333\alpha}^2W_0W_{0,\alpha})\nonumber\\
&+&{\frac {\epsilon^2}{6}}(6a_{3\tau\alpha3\eta\xi\lambda\kappa}^3
U_{1\alpha}U_{0\xi,\eta}U_{0\kappa,\lambda}
+12a_{3\tau\alpha333\lambda\kappa}^3
U_{1\alpha}W_0U_{0\kappa,\lambda} +6a_{3\tau\alpha33333}^3
U_{1\alpha}W_0^2\nonumber\\
&&+3a_{3\tau3\alpha\eta\xi\lambda\kappa}^3
W_{0,\alpha}U_{0\xi,\eta}U_{0\kappa,\lambda}
+6a_{3\tau3\alpha\eta\xi33}^3 W_{0,\alpha}U_{0\xi,\eta}W_0
+3a_{3\tau3\alpha3333}^3 W_{0,\alpha}W_0^2)\nonumber\\
&+&{\cal
O}(\epsilon^3,\nu_1\epsilon)=0~~~~~~(\tau=1,2),\label{e4.6}
\end{eqnarray}
\begin{eqnarray}
&&a_{33\beta\alpha}^1U_{0\alpha,\beta}+a_{3333}^1W_0
+\nu_1(a_{33\beta\alpha}^1U_{1\alpha,\beta}+3a_{3333}^1W_1)\nonumber\\
&+&{\frac
{\epsilon}{2}}(a_{33\alpha\beta\eta\xi}^2U_{0\xi,\eta}U_{0\beta,\alpha}
+2a_{33\alpha\beta33}^2W_0U_{0\beta,\alpha}
+a_{333333}^2W_0^2)\nonumber\\
&+&{\frac {\epsilon^2}{6}}(a_{33\alpha\beta\eta\xi\lambda\kappa}^3
U_{0\beta,\alpha}U_{0\xi,\eta}U_{0\kappa,\lambda}
+3a_{33\alpha\beta\eta\xi33}^3 U_{0\beta,\alpha}U_{0\xi,\eta}W_0\nonumber\\
&&+3a_{33\alpha\beta3333}^3 U_{0\beta,\alpha}W_0^2 +a_{33333333}^3
W_0^3)\nonumber\\
&+&{\cal O }(\epsilon^3,\nu_1\epsilon)=0.\label{e4.7}
\end{eqnarray}
 It should be noted that boundary conditions
(\ref{e4.6}) and (\ref{e4.7}) can be expanded to any needed order.
But, for the problem we consider here, we omit ${\cal O
}(\epsilon^3,\nu_1\epsilon)$  and higher-order terms, since the
purpose is to deduce the leading-order behavior. We note that the
above three equations contain eight unknowns $U_{0\alpha}$,
$W_{0}$, $U_{1\alpha}$, $W_{1}$ and $U_{2\alpha}$. Thus we need
another five equations to have a closed system.

\ind Substituting Eq. (\ref{e4.2}) into (\ref{e3.5}), the
left-hand side becomes a series in $s$,  and all the coefficients
of $s^n (n=0,1,2,\cdots)$ should be zero. As a result, we have two
sets ($\tau=1,2$) of infinitely many equations. Among them, we
only consider those contain the eight unknowns as in (\ref{e4.6})
and (\ref{e4.7}). Actually, from the coefficients of $s^0$ and
$s^1$, we obtain
\begin{eqnarray}
&&a_{\alpha\tau\gamma\beta}^1U_{0\beta,\alpha\gamma}
+2a_{3\tau3\alpha}^1U_{1\alpha}
+(a_{3\tau\alpha3}^1+a_{\alpha\tau33}^1)W_{0,\alpha}\nonumber\\
&+&\epsilon
(a_{\alpha\tau\gamma\beta\eta\xi}^2U_{0\beta,\alpha\gamma}U_{0\xi,\eta}
+2a_{3\tau\alpha3\eta\xi}^2U_{0\xi,\eta}U_{1\alpha}
+a_{\alpha\tau\gamma\beta33}^2U_{0\beta,\alpha\gamma}W_0
\nonumber\\
&&+(a_{3\tau3\beta\eta\xi}^2+a_{\beta\tau33\eta\xi}^2)U_{0\xi,\eta}W_{0,\beta}
+2a_{3\tau\alpha333}^2U_{1\alpha}W_0\nonumber\\
&&+(a_{3\tau3\beta33}^2+a_{\beta\tau3333}^2)W_{0}W_{0,\beta})\nonumber\\
&+& {\epsilon^2}(H_5)=0~~~~~~(\tau=1,2),\label{e4.3}
\end{eqnarray}

\begin{eqnarray}
&&a_{\alpha\tau\gamma\beta}^1U_{1\beta,\alpha\gamma}
+12a_{3\tau3\alpha}^1U_{2\alpha}
+3(a_{3\tau\alpha3}^1+a_{\alpha\tau33}^1)W_{1,\alpha}\nonumber\\
&+&\epsilon
(a_{\beta\tau\gamma\alpha\eta\xi}^2(U_{1\alpha,\beta\gamma}U_{0\xi,\eta}
+U_{0\alpha,\beta\gamma}U_{1\xi,\eta})
+4(a_{\beta\tau3\alpha\eta3}^2+a_{3\tau\beta\alpha\eta3}^2)
U_{1\alpha,\beta}U_{1\eta}\nonumber\\
&&+12a_{3\tau\alpha3\eta\xi}^2U_{0\xi,\eta}U_{2\alpha}
+a_{\beta\tau\gamma\alpha33}^2U_{1\alpha,\beta\gamma}W_0
+(3a_{3\tau3\beta\eta\xi}^2+a_{\beta\tau33\eta\xi}^2)
U_{1\xi,\eta}W_{0,\beta}\nonumber\\
&& +12a_{\alpha\tau3333}^2U_{2\alpha}W_0
+3(a_{3\tau3\beta33}^2+a_{\beta\tau3333}^2)
(W_{0}W_{1,\beta}+W_{1}W_{0,\beta})\nonumber\\
&&+2a_{3\tau\alpha3\eta\xi}^2U_{1\xi,\eta}U_{1\alpha}
+2a_{\alpha\tau\beta3\eta3}^2U_{1\eta}W_{0,\alpha\beta}
+a_{\alpha\tau3\beta3\eta}^2W_{0,\alpha\beta}W_{0,\eta}\nonumber\\
&&+3a_{\alpha\tau\eta\xi33}^2U_{0\xi,\alpha\eta}W_1
+18a_{3\tau\alpha333}^2U_{1\alpha}W_1
+(9a_{3\tau3\alpha33}^2+3a_{\alpha\tau3333}^2)W_{0,\alpha}W_1\nonumber\\
&&+3(a_{3\tau3\alpha\eta\xi}^2+a_{\alpha\tau33\eta\xi}^2)
U_{0\xi,\eta}W_{1,\alpha})\nonumber\\
&+&{\epsilon^2}(H_6)=0 ~~~~~~~(\tau=1,2). \label{e4.4}
\end{eqnarray}
Similarly, substituting Eq. (\ref{e4.2}) into (\ref{e3.6}), we
have a set of infinitely-many equations. We only use the equation
coming from the coefficient of $s^0$ since only it contains the
eight unknowns mentioned before. The equation takes the form:
\begin{eqnarray}
&&2(a_{33\beta\alpha}^1+a_{\alpha3\beta3}^1)U_{1\alpha,\beta}
+a_{\alpha3\beta3}^1W_{0,\alpha\beta}+6a_{3333}^1W_1\nonumber\\
&+& \epsilon(
2(a_{\alpha3\beta3\eta\xi}^2+a_{33\beta\alpha\eta\xi}^2)
U_{1\alpha,\beta}U_{0\xi,\eta}
+2(a_{\alpha3\xi333}^2+a_{33\xi\alpha33}^2)U_{1\alpha,\xi}W_0\nonumber\\
&&+2a_{\beta3\alpha\eta\xi3}^2U_{0\alpha,\eta\beta}U_{1\xi}
+4a_{333\alpha\xi3}^2W_{0,\alpha}U_{1\xi}
+4a_{\alpha333\xi3}^2U_{1\alpha}U_{1\xi}\nonumber\\
&&+a_{\alpha3\beta3\eta\xi}^2W_{0,\alpha\beta}U_{0\xi,\eta}
+a_{\beta3\gamma\alpha3\xi}^2U_{0\alpha,\beta\gamma}W_{0,\xi}
+a_{\alpha3\beta333}^2W_{0,\alpha\beta}W_0\nonumber\\
&& +2a_{\alpha3333\xi}^2(U_{1\alpha} +W_{0,\alpha})W_{0,\xi}
+6a_{3333\alpha\xi}^2U_{0\xi,\alpha}W_1 +6a_{333333}^2W_0W_1)\nonumber\\
&+&{\epsilon^2}(H_7)=0.\label{e4.5}
\end{eqnarray}
Now, the equations (\ref{e4.6}) to (\ref{e4.5}) provide the eight
governing equations for the eight unknowns  $U_{0\alpha}$,
$W_{0}$, $U_{1\alpha}$, $W_{1}$ and $U_{2\alpha}$ and we have a
closed system to work with.

\ind To further simplify the two-dimensional system of Eqs.
(\ref{e4.6}) to (\ref{e4.5}), we shall further use the smallness
of the parameter $\epsilon$ through asymptotic expansions. By a
perturbation method, from (\ref{e4.3})  we obtain
\begin{eqnarray}
U_{1\alpha}&=&-{\frac {1}{2A_3}}(a_{\beta\alpha\lambda\gamma}^1
U_{0\lambda,\beta\gamma}+(A_2+A_3)W_{0,\alpha})\nonumber\\
&-&{\frac
{\epsilon}{2A_3^2}}[(A_3(B_2+B_7)-(A_2+A_3)B_4)W_0W_{0,\alpha}\nonumber\\
&& +(A_3
a_{\beta\alpha\gamma\delta33}^2-B_4a_{\beta\alpha\gamma\delta}^1)
U_{0\delta,\beta\gamma}W_0\nonumber\\
&&+(A_3(a_{\delta\alpha33\beta\lambda}^2+a_{3\alpha\delta3\beta\lambda}^2)
-(a_{3\gamma\delta3}^1+a_{\delta\gamma33}^1)
a_{3\alpha\gamma3\beta\lambda}^2)
U_{0\lambda,\beta}W_{0,\delta}\nonumber\\
&&+(A_3a_{\beta\alpha\delta\gamma\xi\eta}^2
-a_{\beta\zeta\gamma\delta}^1 a_{3\alpha3\zeta\xi\eta}^2)
U_{0\delta,\beta\gamma}U_{0\eta,\xi}]\nonumber\\
&+& {\epsilon^2}(H_8), \label{e5.1}
\end{eqnarray}
where $A_i$ and $B_i$ $(i=1,2,\cdots)$ are defined in Appendix A.
Substituting (\ref{e5.1}) into (\ref{e4.5}), we obtain
\begin{eqnarray}
W_1&=&{\frac {A_2}{6A_3}}W_{0,\alpha\alpha}+{\frac
{A_2+A_3}{6A_1A_3}}
a_{\beta\gamma\delta\alpha}^1U_{0\alpha,\beta\gamma\delta}+{\cal O
}(\epsilon), \label{e5.2}
\end{eqnarray}
where we only give the expression of the leading-order term of
$W_1$ since the higher-order terms have no influence on the final
asymptotic equations. Substituting (\ref{e5.1}) and (\ref{e5.2})
into (\ref{e4.4}), we obtain
\begin{eqnarray}
U_{2\alpha}&=&{\frac
{1}{24A_1A_3^2}}(A_1a_{\delta\zeta\gamma\beta}^1a_{\zeta\alpha\eta\xi}^1
U_{0\beta,\gamma\delta\eta\zeta}
-(A_2+A_3)^2a_{\zeta\beta\eta\xi}^1U_{0\xi,\eta\zeta\beta\alpha}
\nonumber\\
&&-A_1(A_2+A_3)(A_2W_{0,\alpha\beta\beta}-a_{\zeta\alpha\eta\xi}^1W_{0,\zeta\eta\xi}))
 +{\cal O}(\epsilon). \label{e5.3}
\end{eqnarray}
Substituting (\ref{e5.1}) to (\ref{e5.3}) into (\ref{e4.6}) and
(\ref{e4.7}), we obtain
\begin{eqnarray}
&&A_3U_{0\tau,\alpha\alpha}+A_2W_{0,\tau}+
(A_2+A_3)U_{0\alpha,\alpha\tau}\nonumber \\
&-&{\frac {\nu_1}{6}}(A_3U_{0\tau,\alpha\alpha\beta\beta}
+(3A_2+2A_3)W_{0,\alpha\alpha\tau}
+ 3(A_2+A_3)U_{0\alpha,\alpha\beta\beta\tau})\nonumber\\
&+&{\frac {\epsilon}{A_3}}((A_3
a_{\beta\tau\gamma\delta33}^2-B_4a_{\beta\tau\gamma\delta}^1)
U_{0\delta,\beta\gamma}W_0
+(A_1(B_7-B_4)+A_3B_2)W_0W_{0,\tau}\nonumber\\
&&+(a_{3\tau\alpha3\xi\eta}^2 a_{\beta\alpha\lambda\gamma}^1
+A_3a_{\beta\tau\lambda\gamma\xi\eta}^2
-a_{\beta\zeta\gamma\lambda}^1 a_{3\tau\zeta3\xi\eta}^2)
U_{0\lambda,\beta\gamma}U_{0\eta,\xi} \nonumber\\
&&+(A_1(a_{3\tau\alpha3\beta\lambda}^2-a_{3\tau3\alpha\beta\lambda}^2)
-A_3 a_{\alpha\tau33\beta\lambda}^2)
U_{0\lambda,\beta}W_{0,\alpha})\nonumber\\
&&\nonumber\\
&+&{\epsilon^2}(H_9)=0 , \label{e5.4}\\
&&\nonumber\\
&&A_2U_{0\alpha,\alpha}+A_1W_0+{\frac {\nu_1}{2}}(A_1
U_{0\alpha,\alpha\beta\beta}+A_2W_{0,\alpha\alpha})\nonumber\\
&+&{\frac
{\epsilon}{2}}(a_{33\alpha\beta\eta\xi}^2U_{0\xi,\eta}U_{0\beta,\alpha}
+2B_2W_0U_{0\alpha,\alpha}
+B_1W_0^2)\nonumber\\
&+&{\frac {\epsilon^2}{6}}(a_{33\alpha\beta\eta\xi\lambda\kappa}^3
U_{0\beta,\alpha}U_{0\xi,\eta}U_{0\kappa,\lambda}
+3a_{33\alpha\beta\eta\xi33}^3 U_{0\beta,\alpha}U_{0\xi,\eta}W_0\nonumber\\
&&+3C_2 U_{0\alpha,\alpha}W_0^2 +C_1 W_0^3)=0,\label{e5.5}
\end{eqnarray}
where we have omitted the terms higher than  ${\cal O
}(\epsilon^2,\nu_1)$.

\ind For the two side surfaces, we suppose that they are
traction-free. Also, since for a thin plate  they are much smaller
than the top and bottom surfaces, for the latter we need the
boundary conditions to be satisfied at every point (cf.
(\ref{e4.6}) and (\ref{e4.7})) while for the former we only
require the boundary conditions to be satisfied in an average
sense along the thickness.  By integrating the traction-free
boundary conditions $\Sigma_{2i}=0$ at $x_2=\pm \sqrt{\nu_2}$
(where $\nu_2=b^2/l^2$) along the thickness from $0$ to $a$, we
obtain
\begin{eqnarray}
&&A_3(U_{01,2}+U_{02,1}) \nonumber\\
&-&{\frac {\nu_1}{6}}(2(A_2+A_3)(W_0+U_{01,1}+U_{02,2})_{,12}
+A_3(U_{01,2}+U_{02,1})_{,\alpha\alpha})\nonumber\\
&+& {\frac {\epsilon}{2}}
(a_{21\alpha\beta\gamma\delta}^2U_{0\beta,\alpha}U_{0\delta,\gamma}
+2a_{2133\alpha\beta}^2U_{0\beta,\alpha}W_0)\nonumber\\
&+&{\frac
{\epsilon^2}{6}}(a_{21\alpha\beta\gamma\delta\kappa\lambda}^3
U_{0\beta,\alpha}U_{0\delta,\gamma}U_{0\lambda,\kappa} +
3a_{2133\alpha\beta\gamma\delta}^3
U_{0\alpha,\alpha}U_{0\delta,\gamma}W_0 \nonumber\\
&+&3a_{213333\alpha\beta}^3 U_{0\alpha,\alpha} W_0^2))|_{x_2=\pm
\sqrt{\nu_2}}=0,\label{e5.6}
\end{eqnarray}
\begin{eqnarray}
&&2A_3U_{02,2}+A_2(W_0+U_{0\alpha,\alpha})\nonumber\\
&-&{\frac {\nu_1}{6}}(2(A_2+A_3)(W_0+U_{01,1}+U_{02,2})_{,22}
+(A_1U_{02,2}+A_2(W_0+U_{01,1}))_{,\alpha\alpha})\nonumber\\
&+&{\frac
{\epsilon}{2}}(a_{22\alpha\beta\gamma\delta}^2U_{0\beta,\alpha}U_{0\delta,\gamma}
+2a_{22\alpha\beta33}^2U_{0\beta,\alpha}W_0
+B_2W_0^2)\nonumber\\
&+&{\frac
{\epsilon^2}{6}}(a_{22\alpha\beta\gamma\delta\kappa\lambda}^3
U_{0\beta,\alpha}U_{0\delta,\gamma}U_{0\lambda,\kappa}
+3a_{22\alpha\beta\gamma\delta33}^3
U_{0\beta,\alpha}U_{0\delta,\gamma}W_0 \nonumber\\
&&+3a_{22\alpha\beta3333}^3 U_{0\beta,\alpha}W_0^2
+3C_2W_0^3)|_{x_2=\pm\sqrt{\nu_2}}=0. \label{e5.7}
\end{eqnarray}
\begin{eqnarray}
&&A_2W_{0,2}+(A_2+A_3)U_{0\alpha,\alpha 2}+A_3U_{02,\alpha\alpha}\nonumber\\
&-&{\frac {\nu_1}{12}}((3A_2+2A_3)W_{0,\alpha\alpha
2}+3(A_2+A_3)U_{0\alpha,\alpha\beta\beta 2}
+A_3U_{02,\alpha\alpha\beta\beta})\nonumber\\
&+&\epsilon (H_{10})+ {\epsilon^2}(H_{11} )
|_{x_2=\pm\sqrt{\nu_2}}=0. \label{e5.8}
\end{eqnarray}

\ind Eqs. (\ref{e5.4}) and (\ref{e5.5}) are the three
asymptotically-valid governing equations for the three unknowns
 $U_{0\alpha}$ and $W_0$,  among which $U_{0\alpha}$ are the two
 displacement components of
a point in the middle plane and $W_0$ is the normal strain (along
the thickness direction) of that point. Since the two-dimensional
system of Eqs. (\ref{e5.4}) and (\ref{e5.5}) together with the six
boundary conditions  (\ref{e5.6}) to (\ref{e5.8}) are derived from
the three-dimensional field equations, once the solution of this
system is obtained, the three-dimensional displacement field
(thus, also the strain and stress fields) can be easily
calculated.

\bigbreak\noindent{\bf Remark:} It is easy to see that these
governing equations and boundary conditions have two significant
features which are different from the equations for a standard
plane-stress problem. Firstly, the out-plane normal strain $W_0$
is coupled with the in-plane displacement components
$U_{0\alpha}$. And secondly, there are some $\nu_1$ terms and the
orders of the derivatives of these terms are two-order higher than
the other terms, which indicate the influence of the thickness of
the plate. We shall see later that the thickness has an important
influence on the bifurcations. Thus, a model based on a
plane-stress problem may be defective for capturing the
instability phenomena in a thin plate.

\bigbreak

\section{Asymptotic Normal Form Equations for a Thin Strip}
\setcounter{equation}{0}

\ind Now we consider the case that the plate is a thin strip in
terms that both the thickness and  the width of the plate are much
smaller than the length (this is in agreement with of the
experimental setting of Shaw and Kyriakides 1998). Thus, besides
$\nu_1$ being small, $\nu_2$ is also small. As a result,
$-\sqrt{\nu_2}\le x_2\le \sqrt{\nu_2}$ is a small variable. From
Eqs. (\ref{e5.4})-(\ref{e5.8}), it is clear that the unknowns are
functions of the variable $x_1(=:x)$, the small variable
$x_2(=:y)$  and the three small parameters $\epsilon, \nu_1,
\nu_2$, i.e.,
\begin{eqnarray}
U_{0\alpha}=U_{0\alpha}(x,y;\epsilon, \nu_1,\nu_2),~~~~~
W_0=W_0(x,y;\epsilon, \nu_1,\nu_2). \label{e6.0}
\end{eqnarray}

\ind We assume that the unknowns are sufficiently smooth in $y$
and seek the series expansions in the small variable $y$:
\begin{eqnarray}
U_{01}&=&u_0(x)+y^2u_2(x)+y^4u_4(x)+y^6u_6(x)+\cdots,\nonumber\\
&&+\sqrt{\nu_2}~y\cdot(u_1(x)+y^2u_3(x)+y^4u_5(x)+y^6u_7(x)+\cdots),\nonumber\\
U_{02}&=&\sqrt{\nu_2}\cdot(v_0(x)+y^2v_2(x)+y^4v_4(x)+y^6v_6(x)+\cdots)\nonumber\\
&&+y\cdot(v_1(x)+y^2v_3(x)+y^4v_5(x)+y^6v_7(x)+\cdots),\nonumber\\
W_0&=&w_0(x)+y^2w_2(x)+y^4w_4(x)+y^6w_6(x)+\cdots\nonumber\\
&&+\sqrt{\nu_2}~y\cdot(w_1(x)+y^2w_3(x)+y^4w_5(x)+y^6w_7(x)+\cdots),
\label{e6.1}
\end{eqnarray}
where $\sqrt{\nu_2}$ is introduced into the expansions based on
the assumption that the maximum non-dimensional lateral
displacement of a point in the center line of the middle plane is
${\cal O}(\sqrt{\nu_2})$ (i.e., $U_{02}|_{y=0}= {\cal O
}(\sqrt{\nu_2})$), since in the experiment there was only a small
bending (see Shaw and Kyriakides 1998).

\ind Substituting (\ref{e6.1}) into  (\ref{e5.6}) and (\ref{e5.7})
and omitting terms higher than ${\cal O}(\epsilon^2,\nu_2)$ and
with some manipulations, we obtain
\begin{eqnarray}
&&A_3(u_1+ v_{0x}+\nu_2(3u_3+v_{2x}))\nonumber\\
&-& {\frac {\nu_1}{6}}(2A_2(2v_{2x}+w_{1x}+u_{1xx})
+A_3(6u_3+6v_{2x}+2w_{1x}+3u_{1xx}+v_{0xxx}))\nonumber\\
&+&{\epsilon} ( u_{1} (B_4 (u_{0x} + v_{1}) + B_5 w_{0}) + v_{0x}
(B_7 (u_{0x} + v_{1}) + B_8 w_{0}))
\nonumber\\
&+&{\frac {\epsilon^2}{6}}(3 (v_{0x} (C_{12} u_{0x}^2 + C_{12}
v_{1}^2 + 2 C_{15} v_{1} w_{0} + C_{13} w_{0}^2 +
    2 u_{0x} (C_{14} v_{1} + C_{15} w_{0})) \nonumber\\
&&+
  u_{1} (C_{5} u_{0x}^2 + C_{5} v_{1}^2 + 2 C_{8} v_{1} w_{0} + C_{6} w_{0}^2 +
    2 u_{0x} (C_{7} v_{1} + C_{8} w_{0})))
)=0,\nonumber\\
&&\label{e6.11}
\end{eqnarray}
\begin{eqnarray}
&&A_3(2u_2+ v_{1x}+\nu_2(4u_4+v_{3x}))\nonumber\\
&-& {\frac {\nu_1}{6}}(4A_2(3v_{3x}+w_{2x}+u_{2xx})
+A_3(24u_4+18v_{3x}+4w_{2x}+6u_{2xx}+v_{1xxx}))\nonumber\\
&+&{\epsilon}(2 u_{2} (B_4 (u_{0x} + v_{1}) + B_5 w_{0}) + v_{1x}
(B_7 (u_{0x} + v_{1}) + B_8 w_{0}) )
\nonumber\\
&+&{\frac {\epsilon^2}{6}}(3 (v_{1x} (C_{12} u_{0x}^2 + C_{12}
v_{1}^2 + 2 C_{15} v_{1} w_{0} + C_{13} w_{0}^2 +
    2 u_{0x} (C_{14} v_{1} + C_{15} w_{0})) \nonumber\\
&&+
  2 u_{2} (C_{5} u_{0x}^2 + C_{5} v_{1}^2 + 2 C_{8} v_{1} w_{0} + C_{6} w_{0}^2 +
    2 u_{0x} (C_{7} v_{1} + C_{8} w_{0}))))
=0,\nonumber\\
&&\label{e6.12}
\end{eqnarray}
\begin{eqnarray}
&&A_2(u_{0x}+w_0+v_1)+2A_3v_1
+\nu_2(A_2(u_{2x}+w_2)+3(A_2+2A_3)v_3)\nonumber\\
 &-&{\frac {\nu_1}{6}}(6(3A_2+4A_3)v_3+2(3A_2+2A_3)(w_2+u_{2x})
 \nonumber\\
&&+A_1v_{1xx}+A_2(w_{0xx}+u_{0xxx}))\nonumber\\
 &+&{\frac
{\epsilon}{2}} (B_2 u_{0x}^2+ B_1 v_{1}^2 +2 B_2 v_{1} w_{0} + B_2
w_{0}^2 +
  2 u_{0x} (B_2 v_{1} + B_3 w_{0})
)\nonumber\\
&+&{\frac {\epsilon^2}{6}}(C_{2} u_{0x}^3 + C_{1} v_{1}^3 + 3
C_{2} v_{1}^2 w_{0} + 3 C_{3} v_{1} w_{0}^2 +
 C_{2} w_{0}^3 \nonumber\\
&&+ 3 u_{0x}^2 (C_{3} v_{1} + C_{4} w_{0}) +
 3 u_{0x} (C_{2} v_{1}^2 + 2 C_{4} v_{1} w_{0} + C_{4} w_{0}^2)
)=0,\label{e6.13}
\end{eqnarray}
\begin{eqnarray}
&&A_2(u_{1x}+w_1+2 v_2)+4A_3 v_2+\nu_2(A_2(u_{3x}+4 v_4+w_3)+8A_3 v_4)\nonumber\\
 &-&{\frac {\nu_1}{6}}(24(3A_2+4A_3)v_4+6(3A_2+2A_3)(w_3+u_{3x})
 \nonumber\\
&&+2A_1v_{2xx}+A_2(w_{1xx}+u_{1xxx}))\nonumber\\
&+&\epsilon (2 u_{2} (B_4 u_{1} + B_7 v_{0x}) + (B_7 u_{1} + B_4
v_{0x}) v_{1x} +
 2 B_2 v_{2} (u_{0x} + w_{0}) \nonumber\\
&&+ u_{1x} (B_2 u_{0x} + B_3 w_{0}) +
 (B_3 u_{0x} + B_2 w_{0}) w_{1} \nonumber\\
&&+v_{1} (2 B_1 v_{2} + B_2 (u_{1x} + w_{1}))
)\nonumber\\
&+&{\frac {\epsilon^2}{6}}(3 (C_{2} u_{0x}^2 u_{1x} + 4 C_{14}
u_{0x} u_{2} v_{0x} + 2 C_{3} u_{0x}^2 v_{2} +
  2 C_{4} u_{0x} u_{1x} w_{0}  \nonumber\\
&&+ 4 C_{15} u_{2} v_{0x} w_{0}+
  4 C_{4} u_{0x} v_{2} w_{0} + C_{4} u_{1x} w_{0}^2 + 2 C_{3} v_{2} w_{0}^2 \nonumber\\
&& +
  2 v_{0x} v_{1x} (C_{7} u_{0x} + C_{8} w_{0})+
  2 u_{1} (v_{1x} (C_{14} u_{0x} + C_{12} v_{1} + C_{15} w_{0})\nonumber\\
&& +
    2 u_{2} (C_{7} u_{0x} + C_{5} v_{1} + C_{8} w_{0})) + C_{4} u_{0x}^2 w_{1} \nonumber\\
&&+
  2 C_{4} u_{0x} w_{0} w_{1} + C_{2} w_{0}^2 w_{1} +
  2 v_{1} (v_{0x} (2 C_{12} u_{2} + C_{5} v_{1x}) \nonumber\\
&&+ 2 C_{2} v_{2} (u_{0x} + w_{0}) +
    u_{1x} (C_{3} u_{0x} + C_{4} w_{0}) + (C_{4} u_{0x} + C_{3} w_{0}) w_{1})\nonumber\\
&& +
  v_{1}^2 (2 C_{1} v_{2} + C_{2} (u_{1x} + w_{1})))
) =0. \label{e6.14}
\end{eqnarray}
We note that the above four equations contain fourteen unknowns
$u_0 - u_4$, $v_0 - v_4$, $w_0 - w_3$ and to have a closed system,
we need another ten equations.

\ind Substituting (\ref{e6.1}) into Eq. (\ref{e5.4}) for $\tau=1$,
all the coefficients of $y^n$ ($n=1,2,\cdots$) should be zero. So,
we have a set of infinitely-many equations. However, only the
coefficients of $y^0$, $y^1$ and $y^2$ contain the fourteen
unknowns mentioned above, which yield the following three
equations:
\begin{eqnarray}
&&A_2(u_{0xx}+w_{0x}+ v_{1x})+A_3(2u_{0xx}+v_{1x}+2u_2)\nonumber\\
&&-{\frac {\nu_1}{6}}(A_2(3u_{0xxxx}+6u_{2xx}
+3w_{0xxx}+6w_{2x}+18v_{3x}+3v_{1xxx})\nonumber\\
&&+A_3(4u_{0xxxx}+10u_{2xx}+24u_4
+2w_{0xxx}+4w_{2x}+18v_{3x}+3v_{1xxx}))\nonumber\\
&&+{\epsilon}(B_2 u_{0x} v_{1x} + B_7 u_{0x} v_{1x} + B_3 v_{1x}
w_{0} + B_8 v_{1x} w_{0} +
 u_{0xx} (B_1 u_{0x} + B_2 w_{0})\nonumber\\
&& + 2 u_{2} (B_4 (u_{0x} + v_{1}) + B_5 w_{0}) +
 B_2 u_{0x} w_{0x} + B_2 w_{0} w_{0x} \nonumber\\
&&+
 v_{1} (B_2 u_{0xx} + (B_2 + B_7) v_{1x} + B_3 w_{0x})
)\nonumber\\
&&+{\frac {\epsilon^2}{2}}(C_{12} u_{0x}^2 v_{1x} + C_{2} u_{0x}^2
v_{1x} + 2 C_{15} u_{0x} v_{1x} w_{0} \nonumber\\
&&+
 2 C_{4} u_{0x} v_{1x} w_{0} + C_{13} v_{1x} w_{0}^2+ C_{4} v_{1x} w_{0}^2  \nonumber\\
&&+
 u_{0xx} (C_{1} u_{0x}^2 + 2 C_{2} u_{0x} w_{0} + C_{3} w_{0}^2) +
 2 u_{2} (C_{5} u_{0x}^2 + C_{5} v_{1}^2 + 2 C_{8} v_{1} w_{0}  \nonumber\\
&&+ C_{6} w_{0}^2 +
   2 u_{0x} (C_{7} v_{1} + C_{8} w_{0})) + C_{2} u_{0x}^2 w_{0x} +
 2 C_{3} u_{0x} w_{0} w_{0x}  \nonumber\\
&&+ C_{2} w_{0}^2 w_{0x} +
 v_{1}^2 (C_{3} u_{0xx} + (C_{12} + C_{2}) v_{1x} + C_{4} w_{0x}) +
 2 v_{1} (u_{0x} (C_{2} u_{0xx}  \nonumber\\
&&+ (C_{14} + C_{3}) v_{1x} + C_{4} w_{0x}) +
   w_{0} ((C_{15} + C_{4}) v_{1x} + C_{4} (u_{0xx} + w_{0x})))
) =0,\nonumber\\
&&\label{e6.2}\\
&&\nonumber\\
 &&A_2(u_{1xx}+w_{1x}+2 v_{2x})+2A_3(u_{1xx}+
v_{2x}+3u_3)+{\cal O}(\epsilon)=0,\label{e6.3}\\
&&\nonumber\\
&&A_2(u_{2xx}+3v_{3x}+w_{2x})+A_3(2u_{2xx}+3v_{3x}+12u_4)+{\cal
O}(\epsilon)=0.\label{e6.4}
\end{eqnarray}
The ${\cal O}(\epsilon)$ terms in (\ref{e6.3}) and (\ref{e6.4})
will not make contributions to the final results and are thus not
written out. Similarly, substituting (\ref{e6.1}) into Eq.
(\ref{e5.4}) for $\tau=2$ and from the coefficients of $y^0$,
$y^1$ and $y^2$, we obtain
\begin{eqnarray}
&&A_2(u_{1x}+w_1+2 v_2)+A_3(u_{1x}+ v_{0xx}+4 v_2)\nonumber\\
&&-{\frac {\nu_1}{6}}(A_2(3u_{1xxx}+18u_{3x}
+3w_{1xx}+18w_{3}+72v_4+6v_{2xx})\nonumber\\
&&+A_3(3u_{1xxx}+18u_{3x}+2w_{1xx}+12w_{3}+96v_4+10v_{2xx}+v_{0xxxx}))\nonumber\\
&&+{\epsilon}(2 B_2 u_{0x} v_{2} + 2 B_1 v_{1} v_{2} + 2 B_2 v_{2}
w_{0} +
 v_{0xx} (B_4 (u_{0x} + v_{1}) + B_5 w_{0}) \nonumber\\
&&+
 u_{1x} ((B_2 + B_7) (u_{0x} + v_{1}) + (B_3 + B_8) w_{0}) \nonumber\\
&&+
 v_{0x} (B_4 u_{0xx} + 2 B_7 u_{2} + 2 B_4 v_{1x} + B_5 w_{0x})+ B_3 u_{0x} w_{1} \nonumber\\
&&+
 u_{1} (B_7 u_{0xx} + 2 B_4 u_{2} + 2 B_7 v_{1x} + B_8 w_{0x})  +
 B_2 v_{1} w_{1} + B_2 w_{0} w_{1})
\nonumber\\
&&+{\frac {\epsilon^2}{2}}(C_{12} u_{0x}^2 u_{1x} + C_{2} u_{0x}^2
u_{1x} + 2 C_{5} u_{0x} u_{0xx} v_{0x} +
 4 C_{14} u_{0x} u_{2} v_{0x} \nonumber\\
&&+ 4 C_{7} u_{0x} v_{0x} v_{1x} + 2 C_{3} u_{0x}^2 v_{2} +
 2 C_{15} u_{0x} u_{1x} w_{0} + 2 C_{4} u_{0x} u_{1x} w_{0} \nonumber\\
&&+
 2 C_{8} u_{0xx} v_{0x} w_{0} + 4 C_{15} u_{2} v_{0x} w_{0} +
 4 C_{8} v_{0x} v_{1x} w_{0} + 4 C_{4} u_{0x} v_{2} w_{0} \nonumber\\
&& + C_{13} u_{1x} w_{0}^2 +
 C_{4} u_{1x} w_{0}^2 + 2 C_{3} v_{2} w_{0}^2 + C_{4} u_{0x}^2 w_{1}\nonumber\\
&&+
 v_{0xx} (C_{5} u_{0x}^2 + 2 C_{8} u_{0x} w_{0} + C_{6} w_{0}^2) +
 2 C_{8} u_{0x} v_{0x} w_{0x} + 2 C_{6} v_{0x} w_{0} w_{0x} \nonumber\\
&&+
 2 u_{1} (2 ((C_{14} u_{0x}+  C_{12} v_{1}) v_{1x} + C_{15} v_{1x} w_{0}) +
   u_{0xx} (C_{12} u_{0x} + C_{14} v_{1} + C_{15} w_{0}) \nonumber\\
&&+
   2 u_{2} (C_{7} u_{0x} + C_{5} v_{1} + C_{8} w_{0}) + C_{15} u_{0x} w_{0x} +
   C_{15} v_{1} w_{0x} + C_{13} w_{0} w_{0x}) \nonumber\\
&& +
 2 C_{4} u_{0x} w_{0} w_{1} + C_{2} w_{0}^2 w_{1} +
 v_{1}^2 ((C_{12} + C_{2}) u_{1x} + C_{5} v_{0xx} + 2 C_{1} v_{2} + C_{2} w_{1}) \nonumber\\
&&+
 2 v_{1} (2 C_{2} v_{2} (u_{0x} + w_{0}) +
   v_{0x} (C_{7} u_{0xx} + 2 C_{12} u_{2} + 2 C_{5} v_{1x} + C_{8} w_{0x})\nonumber\\
&& +
   w_{0} ((C_{15} + C_{4}) u_{1x} + C_{8} v_{0xx} + C_{3} w_{1}) \nonumber\\
&&+
   u_{0x} ((C_{14} + C_{3}) u_{1x} + C_{7} v_{0xx} + C_{4} w_{1}))
)=0,\label{e6.5}\\ &&\nonumber\\
&&2A_2(u_{2x}+w_2+3 v_3)+A_3(2u_{2x}+
v_{1xx}+12 v_3) +{\cal O}(\epsilon)=0,\label{e6.6}\\
&&\nonumber\\
&&3A_2(u_{3x}+w_3+4 v_4)+A_3(3u_{3x}+ v_{2xx}+24 v_4)+{\cal
O}(\epsilon)=0. \label{e6.7}
\end{eqnarray}
Substituting (\ref{e6.1}) into (\ref{e5.5}) and from the
coefficients of $y^0$, $y^1$, $y^2$ and $y^3$, we obtain
\begin{eqnarray}
&&A_2(u_{0x}+w_{0}+ v_{1})+2A_3w_0\nonumber\\
&&-{\frac {\nu_1}{6}}(A_2(3u_{0xxx}+6u_{2x}
+3w_{0xx}+6w_{2}+18v_3+3v_{1xx})\nonumber\\
&& +A_3(6u_{0xxx}+12u_{2x}+18v_3+6v_{1xx}))\nonumber\\
&&+{\frac {\epsilon}{2}} (B_2 u_{0x}^2 + B_2 v_{1}^2 + 2 B_2 v_{1}
w_{0} + B_1 w_{0}^2 +
  2 u_{0x} (B_3 v_{1} + B_2 w_{0})
)\nonumber\\
&&+{\frac {\epsilon^2}{6}}(C_{2} u_{0x}^3 + C_{2} v_{1}^3 + 3
C_{3} v_{1}^2 w_{0} + 3 C_{2} v_{1} w_{0}^2 +
 C_{1} w_{0}^3 \nonumber\\
&&+ 3 u_{0x}^2 (C_{4} v_{1} + C_{3} w_{0}) +
 3 u_{0x} (C_{4} v_{1}^2 + 2 C_{4} v_{1} w_{0} + C_{2} w_{0}^2)
) =0, \label{e6.8}\\
&&\nonumber\\
&&A_2(u_{1x}+w_{1}+2 v_{2})+2A_3w_1\nonumber\\
&&-{\frac {\nu_1}{6}}(A_2(3u_{1xxx}+18u_{3x}
+3w_{1xx}+18w_{3}+72v_4+6v_{2xx})\nonumber\\
&&+A_3(6u_{1xxx}+36u_{3x}+144v_4+12v_{2xx}))\nonumber\\
&&+{\epsilon}( 2 u_{2} (B_5 u_{1} + B_8 v_{0x}) + (B_8 u_{1} + B_5
v_{0x}) v_{1x} +
 B_2 u_{1x} (u_{0x} + w_{0}) \nonumber\\
&&+ 2 v_{2} (B_3 u_{0x} + B_2 w_{0})+
 (B_2 u_{0x} + B_1 w_{0}) w_{1} + v_{1} (B_3 u_{1x} + 2 B_2 v_{2} + B_2 w_{1})
)
\nonumber\\
&&+ {\frac {\epsilon^2}{6}}(3 (C_{2} u_{0x}^2 u_{1x} + 4 C_{15}
u_{0x} u_{2} v_{0x}  +
  2 C_{3} u_{0x} u_{1x} w_{0} + 4 C_{13} u_{2} v_{0x} w_{0} \nonumber\\
&&+ 2 C_{4} u_{0x}^2 v_{2}+
  4 C_{4} u_{0x} v_{2} w_{0} + C_{2}( u_{1x} + 2  v_{2} )w_{0}^2 +
  2 v_{0x} v_{1x} (C_{8} u_{0x} + C_{6} w_{0})\nonumber\\
&& +
  2 u_{1} (v_{1x} (C_{15} (u_{0x} + v_{1}) + C_{13} w_{0}) +
    2 u_{2} (C_{8} (u_{0x} + v_{1}) + C_{6} w_{0})) \nonumber\\
&&+ C_{3} u_{0x}^2 w_{1} +
  2 C_{2} u_{0x} w_{0} w_{1} + C_{1} w_{0}^2 w_{1} +
  v_{1}^2 (C_{4} u_{1x} + 2 C_{2} v_{2} + C_{3} w_{1}) \nonumber\\
&&+
  2 v_{1} (v_{0x} (2 C_{15} u_{2} + C_{8} v_{1x}) + C_{4} u_{1x} (u_{0x} + w_{0}) +
    2 v_{2} (C_{4} u_{0x} + C_{3} w_{0}) \nonumber\\
&&+ (C_{4} u_{0x} + C_{2} w_{0}) w_{1})))=0, \label{e6.9}\\
&&\nonumber\\
&&A_2(u_{2x}+w_{2}+3 v_{3})+2A_3w_2+{\cal
O}(\epsilon)=0,\label{e6.10}\\
&&\nonumber\\
&&A_2(u_{3x}+w_{3}+4 v_{4})+2A_3w_3+{\cal
O}(\epsilon)=0.\label{e6.10a}
\end{eqnarray}

\ind Now, we have a closed system: the fourteen equations
(\ref{e6.11}) to (\ref{e6.10a}) provide the governing equations
for fourteen unknowns $u_0$ - $u_4$, $v_0$ - $v_4$ and $w_0$ -
$w_3$. Please note that the boundary conditions (\ref{e5.8}) can
be satisfied automatically, and so it will not be used. It is
still formidable to analyze the solutions of this system directly.
Next, we shall use the smallness of the three parameters
$\epsilon, \nu_1$ and $\nu_2$ to proceed further.

\ind By a regular perturbation expansion, from (\ref{e6.11}) we
can obtain the expression of $u_1$ as a function of $u_0$ and
other unknowns. By substituting this expression of $u_1$ into the
remain equations of (\ref{e6.11}) to (\ref{e6.10a}), we can
eliminate $u_1$ from all these equations. Similarly, by solving
the resulting equation of (\ref{e6.2})$\times A_2$ $+$
(\ref{e6.12})$\times (7A_2+4A_3)$, we can obtain $u_2$ as a
function of $u_0$, $v_0$, $\dots$. And by substituting this
expression of $u_2$ into the remain equations of (\ref{e6.11}) to
(\ref{e6.10a}), we can eliminate $u_2$ from all these equations.
Similarly, we can express $u_1$-$u_4$, $w_0$-$w_3$ and $v_1$-$v_4$
in terms of $u_0$ and $v_0$. The concrete forms are
\begin{eqnarray}
u_1&=&-v_{0x}+{\frac
{A_2\nu_1-3(3A_2+2A_3)\nu_2}{6(A_2+A_3)}}v_{0xxx}\nonumber\\
&+&\epsilon{\frac {3A_2
+2A_3}{2(A_2+A_3)}}u_{0x}v_{0x}+\epsilon^2(H_{12})u_{0x}^2v_{0x},\label{u1}\\
u_2&=&{\frac {A_2^2}{8(A_2+A_3)(2A_2+A_3)}}u_{0xx}\nonumber\\
&-& {\frac {A_2(3A_2+2A_3)}{192(A_2+A_3)^2(2A_2+A_3)}}((15A_2+8
A_3)\nu_1-3(3A_2+2A_3)\nu_2)u_{0xxxx}
\nonumber\\
&+&{\frac {\epsilon(3A_2+2A_3)}{32A_3(A_2+A_3)^3(2A_2+A_3)}}
(4A_3^3B_2-4A_2^3(A_3+B_6)\nonumber\\
&&- 2A_2A_3^2(4B_6-B_2+B_3)-A_2^2A_3(5A_3+4(B_6-B_2+B_3)))
u_{0x}u_{0xx}\nonumber\\
&+&\epsilon^2(H_{13})u_{0x}^2u_{0xx},\label{u2}\\
u_3&=&{\frac {5A_2+4A_3}{12(A_2+A_3)}}v_{0xxx} +{\cal
O}(\epsilon), \label{u3}\\
u_4&=&-{\frac
{A_2(19A_2^2+20A_2A_3+4A_3^2)}{384(A_2+A_3)^2(2A_2+A_3)}}u_{0xxxx}
+{\cal O}(\epsilon),\label{u4}
\end{eqnarray}
\begin{eqnarray}
w_3&=&-{\frac {A_2}{12(A_2+A_3)}}v_{0xxxx}+{\cal
O}(\epsilon),\label{w3}\\
w_2&=&-{\frac
{A_2^2(5A_2+2A_3)}{32(A_2+2A_3)^2(2A_2+A_3)}}u_{0xxx}+{\cal
O}(\epsilon),\label{w2}\\
w_1&=&{\frac {A_2}{2(A_2+A_3)}}v_{0xx} +{\frac
{A_2(5(3A_2+2A_3)\nu_2-A_2\nu_1)}{24(A_2+A_3)^2}}v_{0xxxx}\nonumber\\
&-&\epsilon{\frac
{A_2(19A_2^2+26A_2A_3+82A_3^2)}{16(A_2+A_3)^2(2A_2+A_3)}}u_{0xx}v_{0x} \nonumber\\
&+&\epsilon({\frac
{4A_3^2B_2+A_2^2(B_1+3B_2-4B_3)+4A_2A_3(B_2-B_3)}{8(A_2+A_3)^3}}
\nonumber\\
&&-{\frac {A_2(3A_2+2A_3)}{4(A_2+A_3)^2}})u_{0x}v_{0xx}
\nonumber\\
&+&\epsilon^2(H_{14}u_{0xx}v_{0x}+H_{15}u_{0x}v_{0xx})u_{0x},\label{w1}\\
w_0&=&-{\frac {A_2}{2(A_2+A_3)}}u_{0x}\nonumber\\
&+&{\frac
{3A_2^2(3A_2+2A_3)\nu_2-(71A_2^3+208A_2^2A_3+180A_2A_3^2+48A_3^3)\nu_1}
{96(A_2+A_3)^2(2A_2+A_3)}}u_{0xxx}\nonumber\\
&-&\epsilon{\frac
{4A_3^2B_2+A_2^2(B_1+3B_2-4B_3)+4A_2A_3(B_2-B_3)}{16(A_2+A_3)^3}}u_{0x}^2
\nonumber\\
&+&\epsilon^2(H_{16})u_{0x}^3,\label{w0}
\end{eqnarray}
\begin{eqnarray}
v_4&=&-{\frac {2A_2+A_3}{24(A_2+A_3)}}v_{0xxxx}+{\cal
O}(\epsilon),\label{v4}\\
v_3&=&{\frac
{A_2(A_2^2+8A_2A_3+4A_3^2)}{96(A_2+A_3)^2(2A_2+A_3)}}u_{0xxx}
+{\cal O}(\epsilon),\label{v3}\\
v_2&=&{\frac {A_2}{4(A_2+A_3)}}v_{0xx}+{\frac
{(3A_2+2A_3)((7A_2+2A_3)\nu_2-A_2\nu_1)}{48(A_2+A_3)^2}}v_{0xxxx}\nonumber\\
&-&\epsilon{\frac
{A_2(3A_2+2A_3)^2}{32(A_2+A_3)^2(2A_2+A_3)}}u_{0xx}v_{0x} \nonumber\\
&+&\epsilon({\frac
{4A_3^2B_2+A_2^2(B_1+3B_2-4B_3)+4A_2A_3(B_2-B_3)}
{16(A_2+A_3)^3}}\nonumber\\
&&-{\frac {A_2(3A_2+2A_3)}{8(A_2+A_3)^2}})u_{0x}v_{0xx}
\nonumber\\
&+&\epsilon^2(H_{17}u_{0xx}v_{0x}+H_{18}u_{0x}v_{0xx})u_{0x},\label{v2}\\
v_1&=&-{\frac {A_2}{2(A_2+A_3)}}u_{0x}\nonumber\\
&+&{\frac
{A_2((71A_2^2+98A_2A_3+32A_3^2)\nu_1-3(A_2+2A_3)(3A_2+2A_3)\nu_2)}
{96(A_2+A_3)^2(2A_2+A_3)}}u_{0xxx} \nonumber\\
&-&\epsilon{\frac
{4A_3^2B_2+A_2^2(B_1+3B_2-4B_3)+4A_2A_3(B_2-B_3)}{16(A_2+A_3)^3}}u_{0x}^2
\nonumber\\
&+&\epsilon^2(H_{19})u_{0x}^3.\label{v1}
\end{eqnarray}
By inserting (\ref{u1}) to (\ref{v1}) into (\ref{e6.2}) and
(\ref{e6.5}), we obtain two governing equations for the two basic
unknowns $u_0$ and $v_0$:
\begin{eqnarray}
 u_{0xx}+2D_1\epsilon u_{0x}u_{0xx}+3D_2\epsilon^2 u_{0x}^2u_{0xx} -
 {\frac {\nu}{4}} u_{0xxxx}=0,\label{u0}
\end{eqnarray}
\begin{eqnarray}
&&E ((\epsilon u_{0x} +(D_1-1)\epsilon^2 u_{0x}^2)v_{0x})_x-{\frac
{E \nu_2}{3}}v_{0xxxx}=0, \label{v0}
\end{eqnarray}
where
\begin{eqnarray}
E={\frac {A_3(3A_2+2A_3)}{A_2+A_3}}\label{E}
\end{eqnarray}
is the Young's modulus, and
\begin{eqnarray}
D_1&=&{\frac {1}{8 A_3 (A_2 + A_3)^2 (3 A_2 + 2 A_3)}}(4 A_3^3 B_1
+ 12 A_2 A_3^2 (B_1 - B_2)\nonumber\\
&&~~~~ + 6 A_2^2 A_3 (2 B_1 - 3 B_2 + B_3) +
   3 A_2^3 (B_1 - 3 B_2 + 2 B_3)), \label{d1}\\
D_2&=&{\frac {1}{96 A_3 (A_2 + A_3)^4 (3 A_2 + 2
A_3)}}\cdot\nonumber\\
&\cdot&((16 A_3^4 (A_3 C_{1}-3 B_2^2 ) + 16 A_2 A_3^3 (-6 B_2^2 +
6 B_2 B_3 + 5 A_3 C_{1} -
     4 A_3 C_{2}) \nonumber\\
&+& 18 A_2^5 (C_{1} - 4 C_{2} + 3 C_{3}) +
   8 A_2^2 A_3^2 (-3 B_1 B_2 - 15 B_2^2 + 24 B_2 B_3 - 6 B_3^2 \nonumber\\
&+& 20 A_3 C_{1} -
     32 A_3 C_{2} + 6 A_3 C_{3} + 6 A_3 C_{4}) +
   8 A_2^3 A_3 (-3 B_1 B_2 - 9 B_2^2 \nonumber\\
&+& 3 B_1 B_3 + 21 B_2 B_3 - 12 B_3^2 + 20 A_3 C_{1} -
     50 A_3 C_{2} + 18 A_3 C_{3} + 12 A_3 C_{4})\nonumber\\
&+&
   A_2^4 (-3 B_1^2 - 18 B_1 B_2 - 27 B_2^2 + 24 B_1 B_3 + 72 B_2 B_3 - 48 B_3^2 +
     82 A_3 C_{1} \nonumber\\
&-& 280 A_3 C_{2} + 150 A_3 C_{3} + 48 A_3 C_{4})),\label{d2}\\
 \nu&=&{\frac
{2(3A_2+2A_3)\nu_1}{3(A_2+A_3)}}.\label{d3}
\end{eqnarray}
It can be seen that $D_1$ and $D_2$ are constants which depend on
material constants and $\nu$ is proportional to the small
parameter $\nu_1$. As $u_1-u_4$, $v_1-v_4$, $w_0-w_3$ are
expressed in terms of $u_{0x}$ and $v_{0x}$, once $u_{0x}$ and
$v_{0x}$ are found and all these quantities can also be found.

\ind Integrating (\ref{u0}) with respect to $x$ once, we obtain
\begin{eqnarray}
u_{0x}+D_1\epsilon u_{0x}^2+D_2\epsilon^2u_{0x}^3-{\frac {\nu}{4}}
u_{0xxx}=C,\label{u0a}
\end{eqnarray}
where $C$ is an integration constant. It is important to find the
physical meaning of $C$, since we aim to investigate the
instability phenomena as the physical parameters vary. For that
purpose, we consider the resultant axial force $T$ acting on the
material cross section that is planar and perpendicular to the
strip axis in the reference configuration, and the formula is
\begin{eqnarray}
T=\int_{-b}^{b} \int_{-a}^a \Sigma_{11} dX_3dX_2. \label{T}
\end{eqnarray}
By using Eqs. (\ref{e2.4}), (4.2), (4.8)-(4.10) and
(\ref{u1})-(\ref{v1}) in (\ref{T}), it is possible to express
$\Sigma_{11}$ in terms of $u_{0x}$ and $v_{0x}$. Then, carrying
out the integration in (\ref{T}), we find that
\begin{eqnarray}
T=4abE\epsilon(u_{0x}+D_1\epsilon
u_{0x}^2+D_2\epsilon^2u_{0x}^3-{\frac {\nu}{4}} u_{0xxx}).
\label{T1}
\end{eqnarray}
 Comparing Eqs. (\ref{u0a}) and (\ref{T1}),
we have $C={\frac {T}{4abE\epsilon}}$. Thus, we can rewrite
(\ref{u0a}) as
\begin{eqnarray}
 \epsilon u_{0x}+D_1(\epsilon u_{0x})^2+D_2(\epsilon
u_{0x})^3-{\frac {\nu}{4}}  \epsilon u_{0xxx}={\frac {T} {4ab
E}}.\label{u0b}
\end{eqnarray}
If we retain the original dimensional variable and  let
$U=u_{0X}=\epsilon u_{0x}$ (where $X=l x=X_1$), we have
\begin{eqnarray}
U+D_1U^2+D_2U^3-{\frac {a_1^2}{4}}  U_{XX}&=&\gamma,\label{U}
\end{eqnarray}
where
\begin{eqnarray}
\gamma={\frac {T}{4abE}}\label{gamma1}
\end{eqnarray}
is the engineering stress and $a_1^2={\frac
{2(3A_2+2A_3)a^2}{3(A_2+A_3)}}$.

 Similarly, integrating (\ref{v0}) with respect to $x$ once,
we obtain
\begin{eqnarray}
&&E [\epsilon u_{0x} +(D_1-1)\epsilon^2 u_{0x}^2]v_{0x}-{\frac {E
\nu_2}{3}}v_{0xxx}=D, \label{v0a}
\end{eqnarray}
where $D$ is an integration constant. To find out the physical
meaning of $D$, we consider the resultant shear force $Q$ acting
on the material cross section that is planar and perpendicular  to
the strip axis in the reference configuration, and the formula is
\begin{eqnarray}
Q=\int_{-b}^{b} \int_{-a}^a \Sigma_{12} dX_3dX_2. \label{Q}
\end{eqnarray}
By using Eqs. (\ref{e2.4}), (4.2), (4.8)-(4.10) and
(\ref{u1})-(\ref{v1}) in (\ref{Q}), it is possible to express
$\Sigma_{12}$ in terms of $u_{0x}$ and $v_{0x}$. Then, carrying
out the integration in (\ref{Q}), we find that
\begin{eqnarray}
Q=4abE\sqrt{\nu_2}\epsilon ((\epsilon u_{0x} +(D_1-1)\epsilon^2
u_{0x}^2)v_{0x}-{\frac { \nu_2}{3}}v_{0xxx}) .\label{Q1}
\end{eqnarray}
Comparing Eqs. (\ref{v0a}) and (\ref{Q1}), we have $D={\frac
{Q}{4ab\sqrt{\nu_2}\epsilon}}$. Thus, we can rewrite (\ref{v0a})
as
\begin{eqnarray}
4ab E( (\epsilon u_{0x} +(D_1-1)\epsilon^2 u_{0x}^2)\epsilon
\sqrt{\nu_2} v_{0x}-{\frac {\nu_2}{3}}\epsilon \sqrt{\nu_2}
v_{0xxx})=Q.\label{v0b}
\end{eqnarray}
If we retain the original dimensional variables and  let
 $V=\epsilon \sqrt{\nu_2}v_{0x}$ \\ $(=$  $u_{2,1}|_{(X_2,X_3)=(0,0)}$,
 where $u_{2,1}$ is defined in (2.2)), we have
\begin{eqnarray}
EA (U +(D_1-1)U^2)V-EJV_{XX}= {Q} , \label{V}
\end{eqnarray}
where $A$ is the area of the  cross section and $J$ is the moment
inertia around the $X_3$-axis of the cross section.

\ind Once $U$ (i.e., $u_{0x}$) and $V$ (i.e., $v_{0x}$) are found
from equations (\ref{U}) and (\ref{V}), all the other physical
quantities can be calculated immediately. Also, since these two
equations are derived in a mathematically consistent manner and
contain all the required terms to yield the leading-term behavior
of the original three-dimensional problem, we call them to be the
asymptotic normal form equations of the governing
three-dimensional nonlinear PDE's (2.5) with the traction-free
boundary conditions on the top/bottom and two side surfaces under
the given end axial resultant $T$ and end shear resultant $Q$.

\bigbreak\noindent{\bf Remarks:} The results established by us can
also provide some useful information on some classical results
obtained before in an ad hoc manner. The details are discussed
below.

\ind In the case that $U$ is constant, and the nonlinear terms in
(\ref{U}) are neglected, we have $U=\gamma$. If the shear force
$Q$ is zero and
 the nonlinear term $U^2$ is neglected, then Eq. (\ref{V})
equation becomes the linearized Euler bucking equation, which was
obtained based on the {\it hypothesis} that the bending moment is
proportional to the curvature. Here, we have derived this
classical equation based on a three-dimensional setting in a
mathematically consistent manner without such a hypothesis.

\ind  We can also obtain the moment $M$ acting on the $X_3$-axis
 of the material cross section. The formula is
\begin{eqnarray}
M=-\int_{-b}^{b} \int_{-a}^a \Sigma_{11} (X_2+u_2)dX_3dX_2.
\label{M}
\end{eqnarray}
where $u_2$ is the displacement component along the $X_2$-axis
direction. By the same procedure as above, carrying out the
integration in (\ref{M}), we find that
\begin{eqnarray}
M=-4ablE\epsilon \sqrt{\nu_2}((\epsilon u_{0x} +D_1\epsilon^2
u_{0x}^2)v_{0}-{\frac { \nu_2}{3}}v_{0xx}). \label{M1}
\end{eqnarray}
If we retain the original dimensional variable and  let ${\cal
V}=l\epsilon \sqrt{\nu_2}v_{0}$ \\  ($=u_2|_{(X_2,X_3)=(0,0)}$),
we have
\begin{eqnarray}
M=EJ{\cal V}_{XX}-EA (U +D_1U^2){\cal V} .\label{M2}
\end{eqnarray}
Clearly, we have
\begin{eqnarray}
V={\cal V}_X.\label{VV}
\end{eqnarray}
Note that if we let $M$ be zero and do not consider the effect of
the nonlinear term $U^2$, equation (\ref{M2}) also becomes the
classical Euler bucking equation.

\ind  It should also be noted that in general that  $M_X\neq -Q$
as can be seen from (\ref{M1}) and (\ref{Q1}). But in the case
that $u_{0x}$ is constant, denoting $\hat{x}$  the deformed
coordinate of the $x$ axis (namely, $\hat{x}=X+u_0$), then by
neglecting the ${\cal O}(\epsilon^4
\nu_2^{1/2},\epsilon^2\nu_2^{3/2})$ terms, we have
\begin{eqnarray}
{\frac {dM}{d\hat{x}}}={\frac {M_X}{1+\epsilon
u_{0x}}}=-Q.\label{MV}
\end{eqnarray}
Thus, here we have deduced the restriction under which the above
classical result for a beam is valid.

\section{Solutions for an Infinitely-long Strip}
\setcounter{equation}{0}

\ind According to the experiments (Shaw and Kyriakides 1998, Sun
et al 2000, Tse and Sun 2000), for the phase transitions in a thin
strip due to tension/extention, there are at least three
instability phenomena: (i) there is a formation of the
transformation fronts (manifested as a neck); (ii) the neighboring
two transformation fronts incline a same angle with the axial
axis; (iii) the transformation front can switch to an orientation
with the opposite angle. We refer (i) as a necking-type
instability, (ii) as a shear instability and (iii) as a
front-orientation instability.

\ind Now, we analyze the asymptotic normal form equations
(\ref{U}) and (\ref{V}) in order to shed insight into the
instability phenomena observed in the experiments.

\ind As mentioned in Section 2, we consider this class of
non-convex strain energy functions such that in a one-dimensional
stress setting with a homogeneous strain state the engineering
stress-strain curve has a local maximum and a local minimum, which
is same as that considered in the classical paper by Ericksen
(1975). To the third-order material nonlinearity, the
stress-strain relation in this setting is provided by
\begin{eqnarray}
U+D_1U^2+D_2U^3&=&\gamma \nonumber
\end{eqnarray}
(i.e., in (\ref{U}) by setting $U_{XX}=0$, since $U$ is
independent of $X$ in a homogeneous strain state).  The
requirement that the $\gamma-U$ curve has a local maximum and
minimum is equivalent to
\begin{eqnarray}
D_1<0,~~~~D_2>0,~~~~3D_2<D_1^2<4D_2.\label{e8.1}
\end{eqnarray}

\ind The peak stress value $\gamma_2$, the valley stress value
$\gamma_1$ and the Maxwell stress value $\gamma_m$ can be
expressed in terms of $D_1$ and $D_2$ (cf. Dai \& Cai (2006)):
\begin{eqnarray}
&&\gamma_1={\frac
{2D_1^3-2(D_1^2-3D_2)^{3/2}-9D_1D_2}{27D_2^2}},\nonumber\\
&&\gamma_2={\frac
{2D_1^3+2(D_1^2-3D_2)^{3/2}-9D_1D_2}{27D_2^2}},\nonumber\\
&&\gamma_m={\frac {2D_1^3-9D_1D_2}{27D_2^2}}.\label{e8.2}
\end{eqnarray}
The first normal form equation (\ref{U}) has the same form as that
derived for a slender cylinder composed of an incompressible
hyperelastic material (see Cai and Dai 2006 and Dai and Cai 2006).
In those two papers, it has been shown that this equation can be
used to describe the necking-type instability and to capture the
main features of the structure response (engineering
stress-strain) curve. Thus, here we shall not study this equation
and discuss the necking-type instability further. Instead, we
concentrate on the other two instabilities: the shear instability
and the front-orientation instability. For that purpose, we shall
conduct a detailed study on the second normal form equation
(\ref{V}).

\ind We focus on the case that there are two transformation
fronts. If they are some distance away from the two ends of the
strip and any other transformation front (if present), without
loss of generality, we can take the strip to be infinitely-long.
Also we consider the case that the resultant shear force is zero
(for an infinitely-long strip, it has to be zero otherwise the
moment is infinite). Then to determine the solutions of the normal
form equation (\ref{V}) becomes an eigenvalue problem with the
eigenvalue equation
\begin{eqnarray}
EA (U +(D_1-1)U^2)V-EJV_{XX}= 0 , \nonumber
\end{eqnarray}
or
\begin{eqnarray}
c^2 (-U +(1-D_1)U^2)V+V_{XX}= 0 , \label{V0}
\end{eqnarray}
where $c=\sqrt{EA/EJ}={\frac {\sqrt{3}} b}$ is a large parameter
for a strip, and the boundary conditions are
\begin{eqnarray}
V= 0 , ~~~~{\rm at} ~~~~ X=\pm \infty. \label{B0}
\end{eqnarray}

\ind To solve (\ref{V0}) under (\ref{B0}), one needs to solve the
first normal form equation  (\ref{U}) to get $U(X;\gamma)$. It can
be viewed that $\gamma$ is the eigenvalue for  (\ref{V0}). Denote
$f(X;\gamma)=-U+(1-D_1)U^2$, then we can write (\ref{V0}) as
\begin{eqnarray}
V_{XX}+c^2 f(X;\gamma)V =0. \label{e7.4}
\end{eqnarray}
Since $c$ is a large parameter, it is possible to use the WKB
method to construct the leading-order asymptotic solution (see
Holmes 1998). There are three cases. If $f(X;\gamma)>0$ for
$-\infty<X<+\infty$, the general solution (to the leading order)
is
\begin{eqnarray}
V=
 E_1{\frac {e^{ic\int \sqrt{f(X;\gamma)}dX}}{[f(X;\gamma)]^{1/4}}}
 + E_2{\frac {e^{-ic\int \sqrt{f(X;\gamma)}dX}}{[f(X;\gamma)]^{1/4}}}.\label{e7.5}
\end{eqnarray}
If $f(X;\gamma)<0$ for $-\infty<X<+\infty$, the general solution
is
\begin{eqnarray}
V=
 E_3{\frac {e^{c\int \sqrt{-f(X;\gamma)}dX}}{[-f(X;\gamma)]^{1/4}}}
 + E_4{\frac {e^{-c\int \sqrt{-f(X;\gamma)}dX}}{[-f(X;\gamma)]^{1/4}}}.\label{e7.6}
\end{eqnarray}
If there are two turning points at $X_0 (>0)$ and $-X_0$, i.e.,
$f(\pm X_0;\gamma)=0$, $f(X;\gamma)<0$ for $X_0<X<+\infty$ and
$-\infty<X<-X_0$ and $f(X;\gamma)>0$ for $-X_0<X<X_0$, the general
solution is
%\begin{eqnarray}
%V=\left\{\begin{array}{ll}
%C_1[-f(X;\gamma)]^{-1/4}e^{-c\int_{X_0}^X \sqrt{-f(t;\gamma)}dt}&{\rm for}~~X>X_0+\delta_1,\\
%~~+C_1'[-f(X;\gamma)]^{-1/4}e^{c\int_{X_0}^X \sqrt{-f(t;\gamma)}dt}&~~~\\
% C_2{\rm Ai}(C_0(X-X_0))+C_3{\rm Bi}(C_0(X-X_0))&{\rm for}~~X\in (X_0-\delta_1,X_0+\delta_1),\\
% C_4[f(X;\gamma)]^{-1/4}\sin(c\int_0^X
%\sqrt{f(t;\gamma)}dt)
%&{\rm for}~~X\in [-X_0+\delta_2,X_0-\delta_1],\\
%~~+C_5[f(X;\gamma)]^{-1/4}\cos(c\int_0^X \sqrt{f(t;\gamma)}dt)
%&~~~\\
% C_6{\rm Ai}(C_0(X+X_0))+C_7{\rm Bi}(C_0(X+X_0))&{\rm for}~~X\in (-X_0-\delta_2,-X_0+\delta_2),\\
%C_8'[-f(X;\gamma)]^{-1/4}e^{-c\int_{-X_0}^X \sqrt{-f(t;\gamma)}dt}&{\rm for}~~X<-X_0-\delta_2,\\
%~~+C_8[-f(X;\gamma)]^{-1/4}e^{c\int_{-X_0}^X
%\sqrt{-f(t;\gamma)}dt}&~~~
%\end{array} \right. \nonumber\\&&\label{e7.7}
%\end{eqnarray}
\begin{eqnarray}
V=\left\{\begin{array}{lr}
(-f(X;\gamma))^{-1/4}(C_1e^{-c\int_{X_0}^X \sqrt{-f(t;\gamma)}dt}
+C_1'e^{c\int_{X_0}^X \sqrt{-f(t;\gamma)}dt})\\
\hspace{8cm}{\rm for}~~X>X_0+\delta_1,\\
 C_2{\rm Ai}(C_0(X-X_0))+C_3{\rm Bi}(C_0(X-X_0))\\
\hspace{6.1cm}{\rm for}~~X\in (X_0-\delta_1,X_0+\delta_1],\\
 (f(X;\gamma))^{-1/4}(C_4\sin(c\int_0^X
\sqrt{f(t;\gamma)}dt)+C_5\cos(c\int_0^X
\sqrt{f(t;\gamma)}dt))\\
\hspace{6cm}{\rm for}~~X\in [-X_0+\delta_2,X_0-\delta_1],\\
 C_6{\rm Ai}(C_0(X+X_0))+C_7{\rm Bi}(C_0(X+X_0))\\
\hspace{5.5cm}{\rm for}~~X\in [-X_0-\delta_2,-X_0+\delta_2),\\
(-f(X;\gamma))^{-1/4}(C_8'e^{-c\int_{-X_0}^X
\sqrt{-f(t;\gamma)}dt}+C_8e^{c\int_{-X_0}^X
\sqrt{-f(t;\gamma)}dt})\\
\hspace{7.8cm}{\rm for}~~X<-X_0-\delta_2,
\end{array} \right. \nonumber\\&&\label{e7.7}
\end{eqnarray}
where $C_1$, $C_1'$, $\cdots$, $C_8$ are arbitrary constants,
${\rm Ai}(\cdot)$ and ${\rm Bi}(\cdot)$ are the Airy functions of
the first and second kinds respectively,
$C_0=|c^2f_X(X_0;\gamma)|^{1/3}$ and $\delta_1$, $\delta_2$ are
quantities of ${\cal O}(c^{-2/3})$.

\ind According to the experiments, the shear instability happens
immediately after the formation of the two transformation fronts.
Thus, we consider the anti-solitary wave solution of the first
normal form equation (\ref{U}) (which corresponding to the profile
with two transformation fronts; cf. Figure 5 of Dai and Cai 2006).
According to (6.9) of Dai and Cai (2006), the solution expression
is
\begin{eqnarray}
U&=&{\frac {g_{2max}-\alpha_2H_2 \tanh^2({\frac {X}{ga_1}})}{1-H_2
\tanh^2({\frac {X}{ga_1}})}},\label{e7.8}
\end{eqnarray}
where
\begin{eqnarray}
H_2={\frac {g_{2max}-g_1}{\alpha_2-g_1}},~~~~~ g={\frac
{\sqrt{2}}{\sqrt{(\alpha_2-g_1)(g_{2max}-g_1)D_2}}},\label{e7.9}
\end{eqnarray}
and $g_1$ is a double root of and $g_{2max}$ and $\alpha_2$ are
simple roots of
\begin{eqnarray}
{\frac 1 2}U^2+{\frac 1 3}D_1U^3+{\frac 1 4}D_2U^4-\gamma
U-H=0.\label{e7.10}
\end{eqnarray}
For this solution, it is easy to deduce that there are two turning
points at $\pm X_0$ given by
\begin{eqnarray}
X_0=ga_1~{\rm arctanh} \sqrt{\frac
{g_{2max}-U_0}{H_2(\alpha_2-U_0)}},\label{e7.11}
\end{eqnarray}
where
\begin{eqnarray}
U_0=U(X_0;\gamma)={\frac 1 {1-D_1}}.\label{e7.12}
\end{eqnarray}
Thus, in this case the general solution is given by (\ref{e7.7}).
Upon using (\ref{B0}), it can be seen that $C_1'=C_8'=0$. Also,
simple calculations show that $f_X(X_0;\gamma)=U_X(X_0;\gamma)$,
and thus
$C_0=|c^2f_X(X_0;\gamma)|^{1/3}=|c^2U_X(X_0;\gamma)|^{1/3}$.

{\bf Remark:} Since $a_1$ is proportional to the strip thickness
$a$ (see the relation below (5.40)), thus the position of the
turning point is also proportional to $a$. Thus, the thickness of
the strip plays an important role in the instability phenomena.

\ind By using the matching conditions at the neighborhood of
$X=X_0$, we can obtain relationships between constants $C_1$,
$C_2$, $C_3$, $C_4$ and $C_5$ in (\ref{e7.7}). At the neighborhood
of $X=X_0$, we have $f(X;\gamma) \sim U_X(X_0;\gamma)(X-X_0)$. The
integral in (\ref{e7.7})$_3$ can be written as
\begin{eqnarray}
c\int_0^X \sqrt{f(t;\gamma)}dt&=&c\int_0^{X_0}
\sqrt{f(t;\gamma)}dt-c\int_X^{X_0} \sqrt{f(t;\gamma)}dt\nonumber\\
&=&f_1(\gamma)-{\frac 2 3}|C_0(X_0-X)|^{3/2},\label{e7a1}
\end{eqnarray}
where
\begin{eqnarray}
f_1(\gamma)&=&c\int_0^{X_0} \sqrt{f(t;\gamma)}dt=a_1c\sqrt{\frac
{1-D_1}{2D_2}}\int_{U_0}^{g_{2max}}
 \sqrt{\frac {U(U-U_0)}{(\alpha_2-U)(g_{2max}-U)}}{\frac {dU}{U-g_1}}\nonumber\\
&=&{\frac {a_1}
b}(e_1\Pi(\beta_1^2,\tilde{\kappa})+\tilde{g}\Pi(\beta_2^2,\tilde{\kappa})+e_2K(\tilde{\kappa}))
,\label{e7a2}
\end{eqnarray}
\begin{eqnarray}
e_1&=&{\frac
{g_1(U_0-g_1)}{(\alpha_2-g_1)(g_{2max}-g_1)}}\tilde{g},~~~
~~~e_2=-{\frac
{\alpha_2(\alpha_2-U_0)}{(\alpha_2-g_{2max})(\alpha_2-g_1)}}\tilde{g},\nonumber\\
\tilde{g}&=&-{\frac
{\sqrt{6}(\alpha_2-g_{2max})}{\sqrt{g_2U_0(\alpha_2-U_0)D_2}}},~~~~~
\tilde{\kappa}=\sqrt{\frac
{\alpha_2(g_{2max}-U_0)}{g_{2max}(\alpha_2-U_0)}},\nonumber\\
\beta_1^2&=&{\rm tanh}^2( {\frac {X_0}{ga_1}}),~~~~
\beta_2^2=H_2\beta_1^2,\label{e7a3}
\end{eqnarray}
$\Pi(\cdot,\cdot)$ and $K(\cdot)$ are the complete elliptic
integrals of the third and the first kind respectively. By
comparing the asymptotic expansions of the second and the third
forms in (\ref{e7.7}), we obtain
\begin{eqnarray}
C_2&=&\sqrt{\frac \pi
2}C_0^{1/4}|U_X(X_0;\gamma)|^{-1/4}[C_4(\sin(f_1(\gamma)-\cos(f_1(\gamma))),\nonumber\\
&&+C_5(\cos(f_1(\gamma))+\sin(f_1(\gamma)))],\nonumber\\
C_3&=&\sqrt{\frac \pi
2}C_0^{1/4}|U_X(X_0;\gamma)|^{-1/4}[C_5(\cos(f_1(\gamma))-\sin(f_1(\gamma))),\nonumber\\
&&+C_4(\cos(f_1(\gamma))+\sin(f_1(\gamma)))].\label{e8.15}
\end{eqnarray}
By comparing the asymptotic expansions of the second and the first
forms in (\ref{e7.7}), we have
\begin{eqnarray}
C_1&=&{\frac 1 2}\pi^{-1/2}C_0^{-1/4}|U_X(X_0;\gamma)|^{1/4}C_2,\nonumber\\
C_3&=&0.\label{e8.17}
\end{eqnarray}
Similarly, by comparing the asymptotic expansions of the fourth
and the third forms in (\ref{e7.7}),  we obtain
\begin{eqnarray}
C_6&=&\sqrt{\frac \pi
2}C_0^{1/4}|U_X(X_0;\gamma)|^{-1/4}[-C_4(\sin(f_1(\gamma)-\cos(f_1(\gamma))),\nonumber\\
&&+C_5(\cos(f_1(\gamma))+\sin(f_1(\gamma)))],\nonumber\\
C_7&=&\sqrt{\frac \pi
2}C_0^{1/4}|U_X(X_0;\gamma)|^{-1/4}[C_5(\cos(f_1(\gamma))-\sin(f_1(\gamma))),\nonumber\\
&&-C_4(\cos(f_1(\gamma))+\sin(f_1(\gamma)))].\label{e8.15a}
\end{eqnarray}
By comparing the asymptotic expansions of the fifth and the fourth
forms in (\ref{e7.7}), we obtain
\begin{eqnarray}
C_8&=&{\frac 1 2}\pi^{-1/2}C_0^{-1/4}|U_X(X_0;\gamma)|^{1/4}C_6,\nonumber\\
C_7&=&0.\label{e8.17a}
\end{eqnarray}
Substituting $(\ref{e8.15})_2$ into $(\ref{e8.17})_2$, we have
\begin{eqnarray}
C_5\cos(f_1(\gamma)+{\frac \pi 4})+C_4\cos(f_1(\gamma)-{\frac \pi
4})=0.\label{e8.18}
\end{eqnarray}
Substituting $(\ref{e8.15a})_2$ into $(\ref{e8.17a})_2$, we have
\begin{eqnarray}
C_5\cos(f_1(\gamma)+{\frac \pi 4})-C_4\cos(f_1(\gamma)-{\frac \pi
4})=0.\label{e8.18a}
\end{eqnarray}
For Eqs. (\ref{e8.18}) and (\ref{e8.18a}) to have nontrivial
solutions, there are two cases. One case is that
\begin{eqnarray}
C_4=0,~~~~f_1(\gamma)=n\pi+\pi/4,
~~~~n=0,1,2,3,\cdots.\label{e8.19}
\end{eqnarray}
It is easy to see from (\ref{e7.7}) that in this case $V$ is
symmetric. Another case is that
\begin{eqnarray}
C_5=0,~~~~f_1(\gamma)=n\pi-\pi/4, ~~~~n=1,2,3,\cdots,\label{e8.20}
\end{eqnarray}
and in this case $V$ is anti-symmetric.

\ind Eqs. (\ref{e8.19}) and (\ref{e8.20}) are the eigenvalue
equations for determining $\gamma$. These two eigenvalue equations
can be rewritten as a uniform expression
\begin{eqnarray}
f_1(\gamma)={\frac {(2N-1)\pi}{4}},
~~~~N=1,2,3,\cdots,\label{e8.21}
\end{eqnarray}
where odd $N$ represents the symmetric solution for $V$ and the
even $N$ represents the anti-symmetric solution.  By substituting
 (\ref{e7a2}) into (\ref{e8.21}), we obtain
\begin{eqnarray}
&&e_1\Pi(\beta_1^2,\tilde{\kappa})+\tilde{g}\Pi(\beta_2^2,\tilde{\kappa})+e_2K(\tilde{\kappa})
={\frac {b} {a_1}}{\frac {(2N-1)\pi}{4}}.\label{e8.24}
\end{eqnarray}

It can be seen that the left hand side of equation (\ref{e8.24})
implicitly depends on $\gamma$ and the right hand side only
depends on the width-thickness ratio and the wave number $N$.
Thus, the width-thickness ratio is a key factor for determining
the stress eigenvalues.

\ind Denote the eigenvalue corresponding to $N$ by $\gamma_{eN}$.
In the case that $D_1=-18$, $D_2=100$ and the Poisson's ratio
($={\frac {A_2}{2(A_2+A_3)}}$) $=1/3$, $a=0.01$ and ${\frac {b}
{a_1}}=10$ (which means $b\approx 0.1333$), we obtain the first
six eigenvalues as follows:
\begin{eqnarray}
\gamma_{e1}&=&\gamma_m+6.045\times 10^{-16},~~~~
\gamma_{e2}=\gamma_m+1.029\times
10^{-41},\nonumber\\
\gamma_{e3}&=&\gamma_m+1.751\times 10^{-67},~~~~
\gamma_{e4}=\gamma_m+2.980\times
10^{-93},\nonumber\\
\gamma_{e5}&=&\gamma_m+5.071\times 10^{-119},~~~~
\gamma_{e6}=\gamma_m+6.830\times 10^{-145} ,\label{e8.25}
\end{eqnarray}
where the Maxwell stress $\gamma_m=0.0168$. We note that these
eigenvalus are very close.

\ind Denoting ${\cal U}$ and ${\cal V}$ the axial and lateral
displacements of a point in the center line  respectively, and
without loss of generality letting the displacement at $(0,0,0)$
be zero, we obtain
\begin{eqnarray}
{\cal U}=\int_0^X U dX,~~~~{\cal V}=\int_0^X V dX. \label{e8.26}
\end{eqnarray}
Curves of $U$, ${\cal U}$, $V$, ${\cal V}$ for $N=1$ and $N=2$ are
plotted in Figure 1. It should be noted that the amplitude of the
eigenfunction $V$ (and then $\cal V$) is undetermined. We have
scaled all the amplitudes to unit for all variable in Figure 2.
\begin{figure}[htb]
\includegraphics{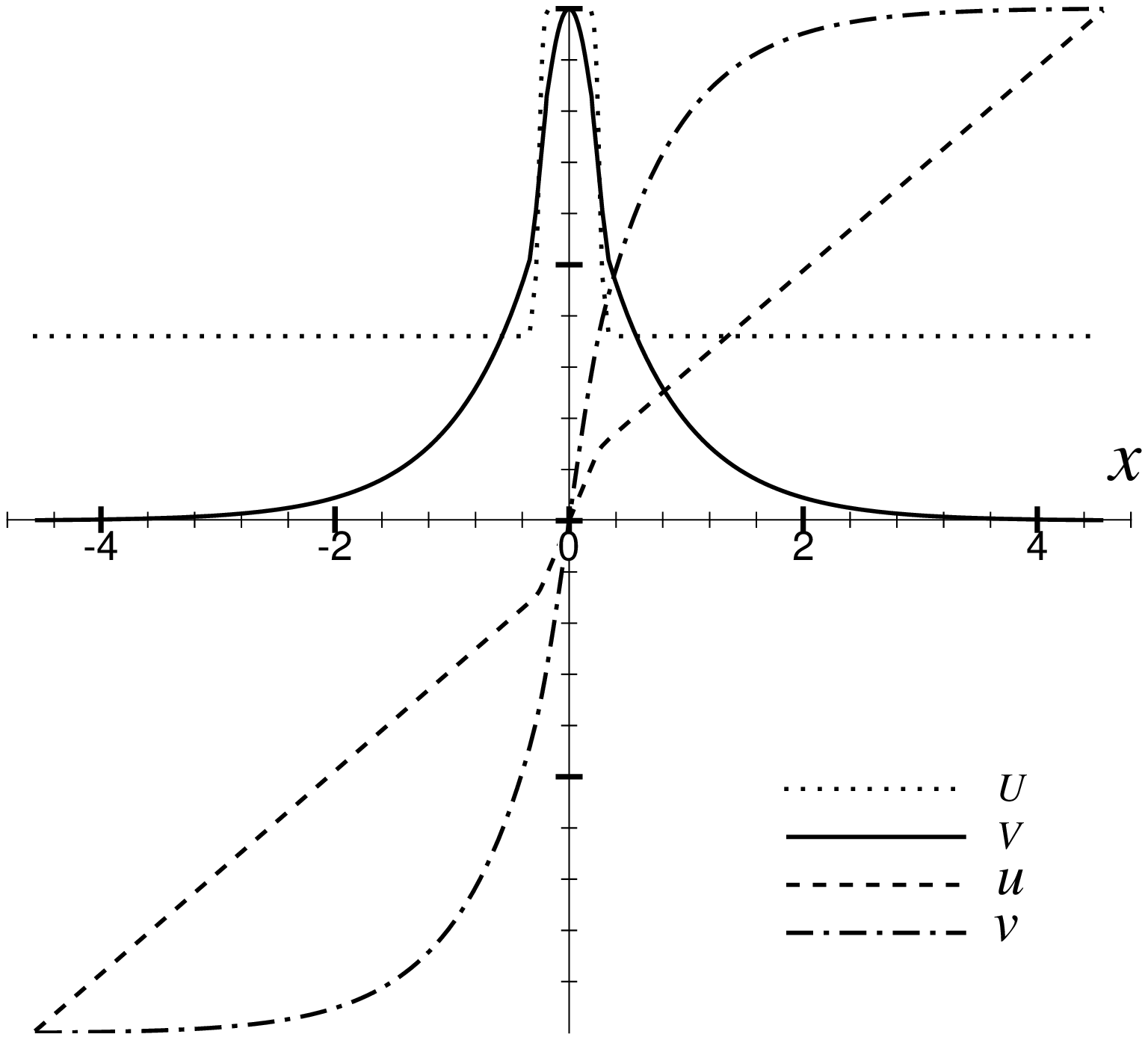} \includegraphics{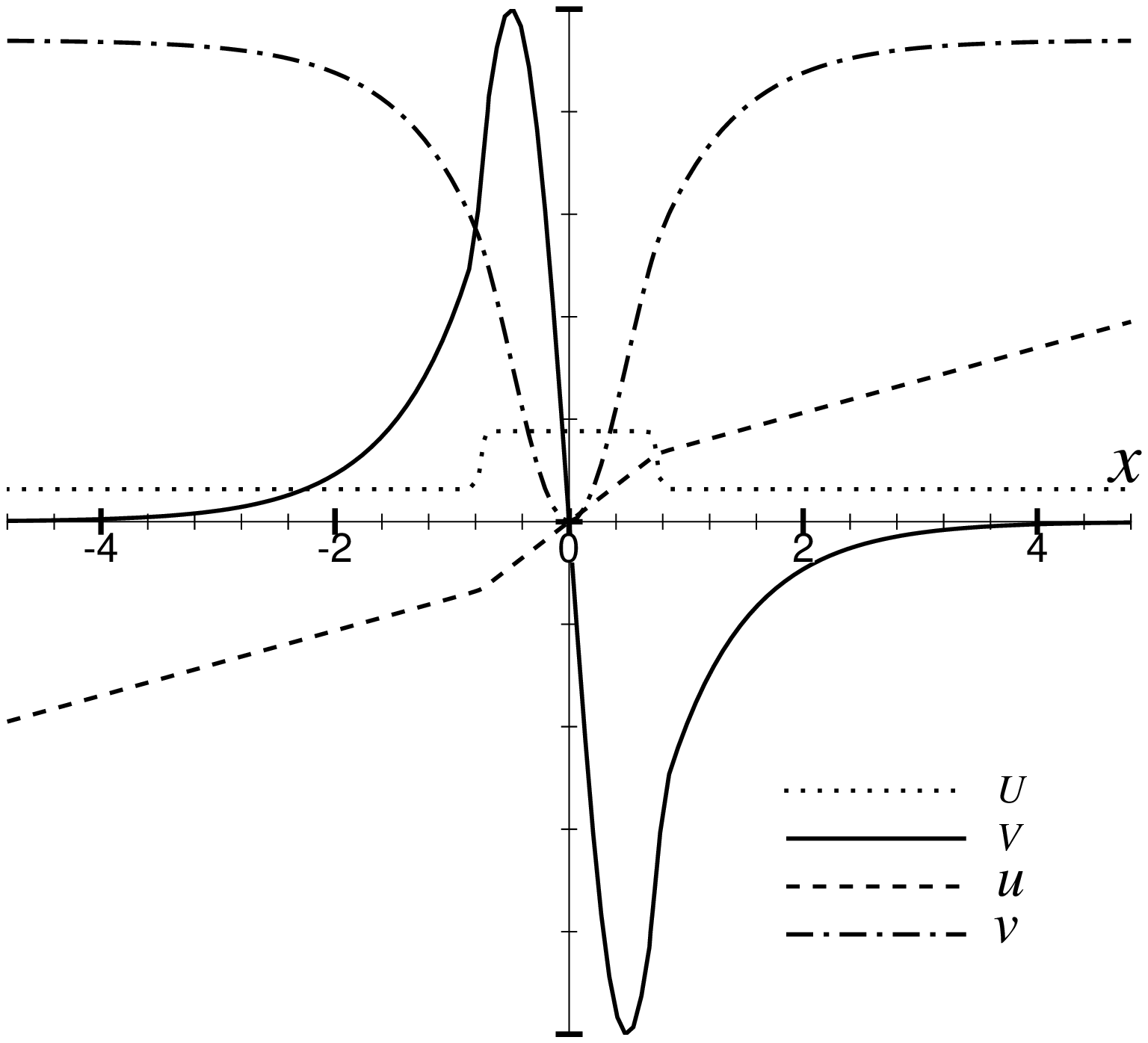} \vspace{5cm} \caption{Curves of $U$,
${\cal U}$, $V$, ${\cal V}$ as $X$ varies. Left: N=1; Right: N=2.
}\label{fig1}
\end{figure}

\ind The  coordinates in the current configuration are (up to the
leading order)
\begin{eqnarray}
x&=&X+{\cal U}-V Y,\nonumber\\
y&=&Y+{\cal V}-\nu U Y,\nonumber\\
z&=&Z-\nu UZ.\label{e8.27}
\end{eqnarray}

The shapes of the thin strip corresponding to $\gamma_{e1}$ to
$\gamma_{e6}$ are plotted in Figure 3. The thickness of the strip
in the current configuration is illustrated by flood contours. It
should be noted that they represent six different modes and do not
need to appear consecutively.
\begin{figure}[htb]
\includegraphics{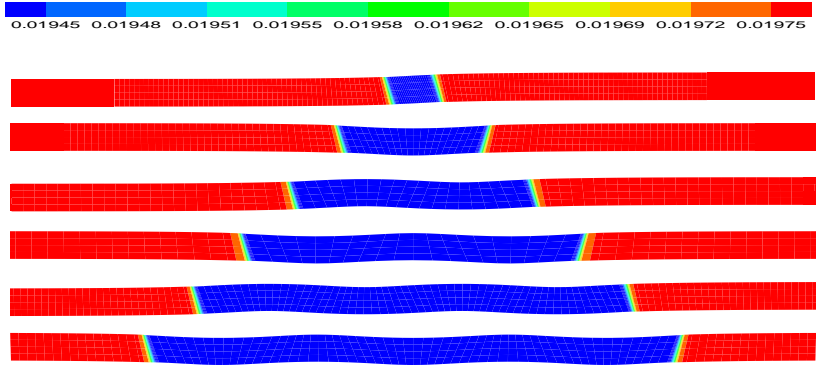} \vspace{6.0cm} \caption{Shapes of the thin strip
corresponding to $\gamma_{e1}$ to $\gamma_{e6}$ (from top to
bottom). The thickness is illustrated by flood
contours.}\label{fig2}
\end{figure}

\ind  From Figure 3, it can be seen that the transformation fronts
are inclined with the strip axis. For $N=1$, the inclined
directions are the same and the position of the turning point
$X_0$ is about $0.2675$ (roughly speaking, $X_0$ represents the
position of the transformation front). For $N=2$, the inclined
directions are opposite and the position of the turning point
$X_0$ is about $0.7619$. In experiments it is found that,
accompanying the formation of two phases, there is a stress drop
($\gamma>\gamma_m$) and the phase fronts become inclined in the
same direction (a shear instability). And then with the developing
of the high-strain phase, the propagating phase front can become
inclined in the opposite direction (orientation instability). Our
analytical results seem to capture these features.

\ind  The results obtained here shows that the inclination of the
transformation front (shear instability) is a phenomenon of
phase-transition-induced buckling. When the phase transition
happens, there is a localized deformation (a neck). Due to that,
the eigenvalue problem (6.5) has a turning point at $X_0$, which
in turn causes the buckling modes. Usually, for a buckling
problem, one often only observes the first mode ($N=1$; the two
transformation fronts are parallel). However, in the present
problem, the second eigenvalue ($N=2$) is very close to the first
eigenvalue. Thus, if there is a slight disturbance, the second
mode may appear. This may explain why the front can switch to an
opposite inclination direction.

We also point out that the solutions obtained above are also valid
for a semi-infinite strip with the boundary conditions:
\begin{eqnarray}
V= 0 , ~~~~~V_X=0, ~~~~{\rm at} ~~~~ X=+\infty,
\end{eqnarray}

if we restrict the spatial interval to $0<X<+\infty$. Then, the
right-half part of Figure 3 describes the different modes of a
single transformation front, which is initially located near the
left end. This corresponds to one experimental situation in Shaw
\& Kyriakides (1998) (see Fig. 1 of that paper).

\section{Conclusions}

In order to better understand several instability phenomena
observed in experiments in a thin SMA strip during the process of
stress-induced phase transitions, we carry out an analytical
study. Mathematically, it is a very challenging problem as one
needs to study the solution bifurcations of nonlinear partial
differential equations in order to capture the instability
phenomena. We start from the formulation of the three-dimensional
field equations. By using the smallness of the thickness and the
maximum strain, through a methodology which combines series
expansions and asymptotic expansions, we derive the
two-dimensional asymptotic equations, which take into account the
lateral deformation and satisfy the traction-free boundary
conditions up to the right order. Then, by further using the
smallness of the width, we derive two one-dimensional asymptotic
normal form equations for the phase transition problem in a thin
strip. These two equations are analyzed and we manage to obtain
some interesting analytic solutions for an infinite long strip
under free-end boundary conditions through the WKB method. Our
analytical results capture several instability phenomena observed
in experiments successfully. It is shown analytically that the
inclination of the transformation front is a phenomenon of
localization-induced buckling (or phase-transition-induced
buckling as the strain localization appears due to the phase
transition). Also, it is demonstrated that there exists a second
mode with a stress eigenvalue very close to that the first one.
Thus, a slight disturbance could cause a switch from the first
mode to the second mode. This, in turn, implies a switch of the
inclination of the transformation front to an opposite direction,
which explains the orientation instability. Our results also
reveal more explicitly the important role played by the thickness
of the strip and show that the width influences the instability
phenomena through the thickness-width ratio rather its magnitude.
In literature, whether the well-known phenomenon of the Luders
band in a mild steel is caused by microscopic effects or
macroscopic effects is still not completely settled issue. Due to
the similarities between stress-induced transformations in a SMA
and the development of Luders band in a mild steel, the present
results also provide a strong mathematical evidence that the
formation of the Luders band is a phenomenon due to macroscopic
effects.

\ind Finally, we point out that as a by-product of the present
study we have also provided a mathematically consistent derivation
of the classical Euler buckling equation, without using the ad hoc
hypothesis that the bending moment is proportional to the
curvature.

\section*{Appendix: Incremental elastic moduli}

\ind For initially isotropic material, in the case that there are
no prestresses, $\Phi$ should be a function of the principle
stretches $\lambda_1$, $\lambda_2$ and $\lambda_3$, namely
$\Phi=\Phi(\lambda_1,\lambda_2, \lambda_3)$. Denote by
$\Phi_j={\frac {\partial \Phi}{\partial
\lambda_j}}|_{\lambda_1=\lambda_2=\lambda_3=1}$,
$\Phi_1=\Phi_2=\Phi_3$ should  vanish since there are no
prestresses. The non-zero first order incremental elastic moduli
can be written as
\begin{eqnarray}
&&A_1=a_{1111}^1=\Phi_{11},\nonumber\\
&&A_2=a_{1122}^1=\Phi_{12},\nonumber\\
&&A_3=a_{1212}^1={\frac {1}{2}}(A_1-A_2),\nonumber\\
&&A_4=a_{1221}^1=A_3,\nonumber
\end{eqnarray}
where $A_2$ and $A_3$ are exactly the Lame's constants for
infinitesimal strain. There are only two independent constants
among $A_i,i=1,2,3,4$.

\ind The non-zero second order incremental elastic moduli can be
written as
\begin{eqnarray}
&&B_1=a_{111111}^2=\Phi_{111},\nonumber\\
&&B_2=a_{111122}^2=\Phi_{112}, \nonumber\\
&&B_3=a_{112233}^2=\Phi_{123}, \nonumber\\
&&B_4=a_{111212}^2={\frac {1}{4}}(2A_2 + 2A_3 + B_1 - B_2),\nonumber\\
&&B_5=a_{331212}^2={\frac {1}{2}}(A_2+B_2-B_3),\nonumber\\
&&B_6=a_{121323}^2={\frac {1}{2}}(B_4-B_5),\nonumber\\
&&B_7=a_{111221}^2=B_4-A_2-A_3 ,\nonumber\\
&&B_8=a_{331221}^2=B_5-A_2,\nonumber\\
&&B_9=a_{123123}^2=B_6-A_3.\nonumber
\end{eqnarray}
There are only three additional independent constants  among
$B_i,i=1\sim 9$.

\ind The non-zero third order incremental elastic moduli can be
written as\begin{eqnarray}
&&C_1=a_{11111111}^3=\Phi_{1111},\nonumber\\
&&C_2=a_{11111122}^3=\Phi_{1112},\nonumber\\
&&C_3=a_{11112222}^3=\Phi_{1122},\nonumber\\
&&C_4=a_{11112233}^3=\Phi_{1123},\nonumber\\
&&C_5=a_{11111212}^3=-{\frac {1}{12}}(6A_2+6A_3-3B_1-3B_2-2C_1+2C_2),\nonumber\\
&&C_6=a_{11112323}^3={\frac {1}{2}}(B_2+C_3-C_4),\nonumber\\
&&C_7=a_{11221212}^3=-{\frac {1}{12}}(6A_2+6A_3-6B_2-C_1-2C_2+3C_3),\nonumber\\
&&C_8=a_{11221313}^3=-{\frac {1}{4}}(A_2-B_2-B_3-C_2+C_4),\nonumber\\
&&C_9=a_{12121212}^3={\frac {1}{8}}(6A_2+6A_3+6B_1-6B_2+C_1-4C_2+3C_3),\nonumber\\
&&C_{10}=a_{12121313}^3={\frac {1}{3}}C_9,\nonumber\\
&&C_{11}=a_{11121323}^3=-{\frac {1}{24}}(3A_1-3B_2+3B_3-C_1+C_2+3C_3-3C_4),\nonumber\\
&&C_{12}=a_{11111221}^3={\frac {1}{12}}(6A_2+6A_3-3B_1-3B_2+2C_1-2C_2),\nonumber\\
&&C_{13}=a_{11112332}^3=-{\frac {1}{2}}(B_2-C_3+C_4),\nonumber\\
&&C_{14}=a_{11221221}^3={\frac {1}{12}}(6A_2+6A_3-6B_2+C_1+2C_2-3C_3),\nonumber\\
&&C_{15}=a_{11221331}^3={\frac {1}{4}}(A_2-B_2-B_3+C_2-C_4),\nonumber\\
&&C_{16}=a_{12121221}^3=-{\frac {1}{8}}(6A_2+6A_3-C_1+4C_2-3C_3),\nonumber\\
&&C_{17}=a_{12211221}^3={\frac {1}{8}}(6A_2+6A_3-2B_1+2B_2+C_1-4C_2+3C_3),\nonumber\\
&&C_{18}=a_{12121331}^3={\frac {1}{3}}C_{16},\nonumber\\
&&C_{19}=a_{12211331}^3={\frac {1}{24}}(6A_1-3B_1-3B_2+6B_3+C_1-4C_2+3C_3),\nonumber\\
&&C_{20}=a_{11123123}^3={\frac {1}{24}}(6A_1+3A_2-3B_1+3B_3+C_1-C_2-3C_3+3C_4),\nonumber\\
&&C_{21}=a_{11123132}^3=C_{20}-{\frac {1}{4}}(A_1+A_2+B_1-B_2),\nonumber\\
&&C_{22}=a_{12122323}^3={\frac
{1}{24}}(6A_2-6A_3+12B_2-12B_3+C_1-4C_2+3C_3).\nonumber
\end{eqnarray}
There are only four additional independent constants  among
$C_i,i=1\sim 22$.

% \appendix

% \section{}
% \label{}

\bigbreak {\bf Acknowledgement}\\
The work described in this paper is supported by a grant from the
Research Grants Council of the HKSAR, China (Project number: CityU
100807) and a grant from the National Natural Science Foundation
of China (Project number: 10672117).

\end{document}